\documentclass[a4paper, 11pt]{article}
\usepackage{amsmath, amssymb, color} %not needed when jheppub is used.
\usepackage{mathrsfs}
\usepackage{slashbox}
\def\hsymbl#1{\smash{\hbox{\huge$#1$}}}
\def\hsymbu#1{\smash{\lower1.7ex\hbox{\huge$#1$}}}
\newcommand{\bs}{\begin{subequations}}
\newcommand{\es}{\end{subequations}}
\newcommand{\bbm}{\begin{bmatrix}}
\newcommand{\ebm}{\end{bmatrix}}
\newcommand{\mrm}{\mathrm}

\newcommand{\p}{\partial}
\newcommand{\diff}{\mathrm{d}}
\newcommand{\iu}{\mathrm{i}}
\newcommand{\e}{{\hspace{.1em}{\mathrm e}}}
\newcommand{\adPhi}{\mathrm{ad}_\Phi}
\newcommand{\adg}{{\mathrm{ad}_{g\Phi}}}
\newcommand{\Xil}{\Xi_\lambda} 
 
\newcommand{\Xmid}{X_{\mathrm{mid}}} 
\newcommand{\Xl}{\mathcal{X}_\lambda} 
\newcommand{\Ginv}{G^{-1}}
\newcommand{\SW}{S^\mathrm{W}}
\newcommand{\calSW}{\boldsymbol{\mathcal{S}}^\mathrm{W}}
\newcommand{\SB}{S^\mathrm{B}}
\newcommand{\calSB}{\boldsymbol{\mathcal{S}}^\mathrm{B}}
\newcommand{\calS}{\boldsymbol{\mathcal{S}}}
\newcommand{\LW}{\Lambda^\mrm{W}}
\newcommand{\Hs}{\mathscr{H}_\mathrm{small}} %small Hilbert space
\newcommand{\tQ}{\widetilde{Q}}
\newcommand{\PsiW}{\Psi^\mathrm{W}}
\newcommand{\bPsiW}{\boldsymbol{\Psi}^\mathrm{W}}
\newcommand{\PhiN}{\Phi^{-}}
\newcommand{\PhiXi}{\Phi^\Xi}
\newcommand{\hatG}{\hat{G}}
\newcommand{\hatGinv}{\hat{G}^{-1}}

\newcommand{\hd}{\widehat{\delta}}
\newcommand{\hdsub}{\widehat{\delta}^\mathrm{sub}}
\newcommand{\xz}{\xi_0}
\newcommand{\ez}{\eta_0}
\newcommand{\calO}{\mathcal{O}}
\newcommand{\hs}{\hspace{.07em}}
\newcommand{\llangle}{\langle\hspace{-0.8mm}\langle}
\newcommand{\bllangle}{\big\langle\hspace{-0.9mm}\big\langle}
\newcommand{\Bllangle}{\Big\langle\hspace{-1.5mm}\Big\langle}
\newcommand{\bgllangle}{\bigg\langle\hspace{-2.2mm}\bigg\langle}
\newcommand{\Bgllangle}{\Bigg\langle\hspace{-2.3mm}\Bigg\langle}
\newcommand{\rrangle}{\rangle\hspace{-0.8mm}\rangle}
\newcommand{\brrangle}{\big\rangle\hspace{-0.9mm}\big\rangle}
\newcommand{\Brrangle}{\Big\rangle\hspace{-1.5mm}\Big\rangle}
\newcommand{\bgrrangle}{\bigg\rangle\hspace{-2.2mm}\bigg\rangle}
\newcommand{\Bgrrangle}{\Bigg\rangle\hspace{-2.3mm}\Bigg\rangle}
\allowdisplaybreaks
\begin{document}
\begin{titlepage}
\rightline{}
% \rightline{\tt arXiv:yymm.nnnn}
\rightline{\tt RIKEN-QHP-197}
\rightline{\tt RIKEN-STAMP-13}
% \rightline{\today}
\begin{center}
\vskip 1.8cm
%title 
{\large \bf{
Relation between the Reducibility Structures 
and between the Master Actions
}}\\
\vskip 0.5cm
{\large \bf{
in the Witten Formulation and the Berkovits Formulation
}}\\
\vskip 0.5cm
{\large \bf{
of Open Superstring Field Theory
}}\\
\vskip 2cm
%author
{\large Yuki {\sc Iimori}$^1$ and Shingo {\sc Torii}$^2$}
\vskip 1cm
%affiliation
{\large 
$^1${\it {Nagoya International Patent Firm,\\ 20-19, Nishiki 1-chome, Naka-ku, Nagoya 460-0003, Japan}}\\[3mm]
$^2${\it {Quantum Hadron Physics Laboratory, RIKEN Nishina Center,\\ Saitama 351-0198, Japan}}\\[5mm]
%emailAdd
yiimori2@gmail.com, shingo.torii@riken.jp
}
\vskip 2.7cm
%abstract
{\large {\bf Abstract}}
\end{center}
\vskip 0.5cm
\baselineskip 16pt

Developing the analysis in JHEP {\bf 03} (2014) 044 [arXiv:1312.1677] by the present authors et al.,
we clarify the relation between the Witten formulation and the Berkovits formulation of open superstring field theory
at the level of the master action, namely the solution to the classical master equation in the Batalin-Vilkovisky formalism,
which is the key for the path-integral quantization.
We first scrutinize the reducibility structure, a detailed gauge structure containing the information about ghost string fields.
Then, extending the condition for partial gauge fixing introduced in the above-mentioned paper to the sector of ghost string fields, 
we investigate the master action.
We show that the reducibility structure and the master action under partial gauge fixing of the Berkovits formulation can be
regarded as the regularized versions of those in the Witten formulation.

\end{titlepage}
%keywords{String Field Theory}
\baselineskip 17pt
\tableofcontents

%-------------------------------------------------------------------------------------------------
\newpage
%%%%%%%%%%%%%%%%%%%%%%%%%%%%%%%%%%%%%%%%%
\section{Introduction}
\label{section:introduction}
\setcounter{equation}{0}
Since the decisive formulation of open bosonic string field theory~\cite{Witten:bosonic},
various attempts have been made to construct manifestly covariant open superstring field theory
based on the Ramond-Neveu-Schwarz formalism~\cite{Witten:super, AMZ, PTY, Berkovits:NS, Berkovits:R, Michishita, Kunitomo, DemocraticSSFT, EKS}.
They can be classified into two types, according to the way to treat the Hilbert space of the superconformal ghost sector
on the world-sheet~\cite{FMS1, FMS2}:
the approaches based on the small Hilbert space, and those based on the large Hilbert space.
The relation between the two types of approach has recently been investigated 
with the technique of partial gauge fixing, especially in the Neveu-Schwarz (NS) sector~\cite{INOT, EOT}.
In the partial gauge fixing, however, ghost string fields, which are necessary for proper gauge fixing, are not taken into consideration;
therefore the scope of the analyses is limited to only a certain aspect of the theories.
In particular, we can say little about the relation from the viewpoint of the path integral.\footnote{
For the case of free theory, the correspondence between the completely gauge-fixed actions in the
Witten formulation and the Berkovits formulation is shown in ref.~\cite{paperI}.
}
The aim of the present paper is to acquire a deeper understanding
by elucidating the relation at the level of the master action,
namely the solution to the classical master equation in the Batalin-Vilkovisky (BV) formalism~\cite{BV1, BV2, Henneaux, GPS},
which is the key for the path-integral quantization of complicated gauge systems such as string field theory.

Historically, the first manifestly covariant open superstring field theory was formulated by Witten~\cite{Witten:super}.
Based on the small Hilbert space, it is a natural extension of open bosonic string field theory. 
However, it has the problem of divergences caused by the picture-changing operator inserted at the string midpoint~\cite{Wendt}.
Among the approaches to overcoming this problem, the Berkovits formulation for the NS sector is remarkable~\cite{Berkovits:NS}.
The theory is constructed without using any picture-changing operators, based on the large Hilbert space.
Recently, the present authors et al.\ have manifested the mechanism of how the problem in the Witten formulation is resolved 
in this formulation~\cite{INOT}.
In the paper~\cite{INOT}, 
we have shown that the action and the gauge transformation in the Berkovits formulation 
can be interpreted as the regularized versions of those in the Witten formulation, 
using the technique of partial gauge fixing.
Imposed on the condition for partial gauge fixing,
the Berkovits action can be written in the form including line integrals of the picture-changing operator, 
rather than its local insertions;
therefore the divergences in the Witten formulation are avoided.
Inspired by this line-integral regularization mechanism, Erler, Konopka, and Sachs have constructed a new open superstring field theory
with the small Hilbert space approach~\cite{EKS}, and its relation to the Berkovits formulation, also,
has been analyzed by the use of partial gauge fixing~\cite{EOT}.\footnote{
For the relation between the Berkovits formulation and the Erler-Konopka-Sachs formulation,
see also ref.~\cite{Erler}.}
In these studies on the relation between the small Hilbert space approach and the large Hilbert space approach, however, 
partial gauge fixing is performed without taking into account ghost string fields,
and an understanding from the viewpoint of the path integral is missing.
In order to deepen our understanding of the relation,
in the present paper we perform a detailed analysis on the NS sector in the Berkovits formulation and 
compare it with the one in the Witten formulation,
extending the condition for partial gauge fixing in ref.~\cite{INOT} to the sector of ghost string fields.
We first scrutinize the detailed gauge structure called reducibility structure.
It contains the information about ghost string fields,
and governs the form of the master action, which is the key for the path-integral quantization in the BV formalism.
We then investigate the relation between the master actions in the two formulations.
Through the analyses, 
we show that the reducibility structure and the master action under partial gauge fixing of the Berkovits formulation can be regarded as 
the regularized versions of those in the Witten formulation.

The present paper is organized as follows.
In section~\ref{section:SSFTs}, we briefly review the Witten formulation and the Berkovits formulation of open superstring field theory,
concentrating on the NS sector.
We then explain in section~\ref{section:Gauge-fixing} partial gauge fixing of the Berkovits formulation introduced in ref.~\cite{INOT}.
After that, extending the condition for partial gauge fixing, we investigate the reducibility structures and the master actions 
in sections~\ref{sec:reducibility} and~\ref{sec:master action}, respectively.
Finally, section~\ref{section:discussion} is allocated for summary and discussion.
Two appendices are provided to supply details of the analyses.

%%%%%%%%%%%%%%%%%%%%%%%%%%%%%%%%%%%%%%%%%
\section{The Witten formulation and the Berkovits formulation}
\label{section:SSFTs}
\setcounter{equation}{0}
In the present section, we review two formulations of open superstring field theory, concentrating on the NS sector. 
One is the Witten formulation~\cite{Witten:super}, and the other is the Berkovits formulation~\cite{Berkovits:NS}.
The action in the former is constructed in the small Hilbert space,
and that in the latter is in the large Hilbert space.
We first summarize the basics of the two Hilbert spaces in subsection~\ref{subsec:small/large};
then we briefly review the Witten formulation in subsection~\ref{Witten formulation}
and the Berkovits formulation in subsection~\ref{Berkovits formulation}.

%----------------------------------------
\subsection{The small Hilbert space and the large Hilbert space}
\label{subsec:small/large}
The small Hilbert space and the large Hilbert space are basic concepts in the Ramond-Neveu-Schwarz formalism.
In order to see the difference between the two spaces, we fermionize the superconformal ghosts $\beta$ and $\gamma$
as in refs.~\cite{FMS1, FMS2}:
\begin{equation} \label{bosonization}
\beta = \e^{-\phi}\p\xi\,,\quad 
\gamma = \eta \e^\phi\,.
\end{equation}
The fields $\xi$ and $\eta$ are fermionic, whereas $\phi$ is bosonic.
Here and in what follows, we omit the normal-ordering symbol with respect to the SL$(2, \mathbb{R})$-invariant vacuum for simplicity,
and use the convention in which appropriate cocycle factors are implicitly included,
so that $\e^{l \phi}$ ($l\in$ odd) anticommute with fermionic operators.
The fundamental operator product expansions (OPEs) of $\xi$, $\eta$, and $\phi$ are given by
\begin{equation}
\xi(z_1)\eta(z_2) \sim \frac{1}{z_1-z_2}\,,\qquad
\phi(z_1)\phi(z_2) \sim -\ln \left(z_1-z_2\right),
\end{equation}
with ``$\sim$'' denoting the equality up to non-singular terms.
In \eqref{bosonization}, the operator $\xi$, whose conformal weight is zero, is accompanied by the derivative symbol $\partial$.
In fact, we can describe superstring theory, not using the bare $\xi$.
In other words, we can describe it without the zero mode $\xz$.\footnote{In the present paper,
an operator $\calO$ of conformal weight $h$ is expanded in the coordinate $z$
on the upper half-plane as
\begin{align}
\calO(z) = \sum_n \frac{\calO_n}{z^{n+h}}\,.
\end{align}}
The superstring Hilbert space containing only the states which can be constructed without $\xz$ is called the \emph{small Hilbert space}, 
and the one including also the states involving $\xz$ is called the \emph{large Hilbert space}.
If a state $A$ is in the small Hilbert space $\Hs$, it is annihilated by the zero mode of $\eta$, and vice versa:
\begin{equation}
A \in \Hs \iff \eta_0 A = 0 \,.
\end{equation}
It follows from the OPEs of $\xi$ and $\eta$ that the zero modes $\xi_0$ and $\eta_0$ satisfy
\begin{equation} \label{xzez}
\xz^2 = \ez^2 = 0\,,\quad \{\xz, \ez\} =1\,.
\end{equation}
Therefore any state $\varphi$ in the large Hilbert space can be written in terms of two states $A$ and $B$ in the small Hilbert space as
\begin{equation} \label{varphi:decomposition} 
\varphi = A +\xi_0 B \,,
\end{equation}
with
\begin{equation}
A = \eta_0 \xi_0 \varphi \,, \quad
B = \eta_0 \varphi \,.
\end{equation}
Thus we could say that the large Hilbert space is twice as large as the small one.

In each of the spaces, there are two important quantum numbers: 
the world-sheet ghost number $\boldsymbol{g}$ and the picture number $\boldsymbol{p}$.
They are defined by the charges $Q_{\boldsymbol{g}}$ and $Q_{\boldsymbol{p}}$ below:
\begin{equation}
Q_{\boldsymbol{g}} = \oint_C \frac{\diff z}{2\pi\iu} \bigl( -bc(z)-\xi\eta(z)\bigr) \,, \qquad
Q_{\boldsymbol{p}} = \oint_C \frac{\diff z}{2\pi\iu} \bigl( -\p\phi(z) +\xi\eta (z)\bigr) \,.
\end{equation}
Here we have used the doubling trick,\footnote{
For the doubling trick, see refs.~\cite{Polchinski:1, Polchinski:2}, for example.
}
and have denoted by $C$ the counterclockwise unit circle 
centered at the origin.
The ghost number and the picture number of the BRST operator $Q$, for example, are one and zero, respectively.
In the fermionized description~\eqref{bosonization}, we have
\begin{equation}
Q = \oint_C \frac{\diff z}{2\pi\iu} \, j_B (z) \,,
\end{equation}
with the BRST current $j_B$ given by\footnote{
The total derivative term $\frac{3}{4}\p^2 c$ makes $j_B$ primary.
}
\begin{equation} \label{j_B}
j_B = c\hs T^\mrm{m} +\eta\e^\phi\hs G^\mrm{m} +bc\p c +\frac{3}{4}\hs\p c\p\phi -\frac{1}{4}\hs c\p^2\phi
-\frac{1}{2}\hs c\p\phi\p\phi -c\eta\p \xi -b\eta\p\eta \e^{2\phi} +\frac{3}{4}\hs\p^2 c \,,
\end{equation}
where $T^\mrm{m}$ is the matter energy-momentum tensor and $G^\mrm{m}$ is the matter supercurrent.
The action of $Q$ upon $\xi$ gives the picture-changing operator~\cite{FMS1, FMS2}
\begin{equation} \label{PCO}
X := Q \cdot \xi  
= \e^\phi G^\mrm{m} +c\p\xi + b\p\eta\e^{2\phi} +\p \bigl( b\eta\e^{2\phi}\bigr)\,,
\end{equation}
which raises picture number by one.
We list the ghost number $\boldsymbol{g}$ and the picture number $\boldsymbol{p}$, 
together with the conformal weight $h$, of various operators in table~\ref{table}.
\begin{table}
\begin{center}
\caption{The ghost number $\boldsymbol{g}$, the picture number $\boldsymbol{p}$, and the conformal weight $h$ of various operators.}
\ \\[-.5ex]
{\renewcommand\arraystretch{1.3}
\begin{tabular}{|c||c|c|c|c|c|c|c|c|c|}
\hline
operator & $b$ & $c$ & $\xi$ & $\eta$ & $\e^{l\phi}$ & $\beta$ & $\gamma$ & $j_B$ & $X$ \\
\hline
$(\boldsymbol{g}, \boldsymbol{p})$ & $(-1,0)$ & $(1,0)$ & $(-1,1)$ & $(1,-1)$ & $(0,l)$ & $(-1,0)$ & $(1,0)$ & $(1,0)$ & $(0,1)$ \\
\hline
$h$ & $2$ & $-1$ & $0$ & $1$ & $-\frac{1}{2}\hs l(l+2)$ & $\frac{3}{2}$ & $-\frac{1}{2}$ & $1$ & $0$ \\
\hline
\end{tabular}
}
\label{table}
\end{center}
\end{table}

Let $\llangle A,B\rrangle$ and $\langle A, B\rangle$ denote 
the Belavin-Polyakov-Zamolodchikov (BPZ) inner product~\cite{BPZ} of states $A$ and $B$ 
in the small Hilbert space and that in the large Hilbert space, respectively.
The inner product in the small Hilbert space, $\llangle A,B\rrangle$, 
vanishes unless the sum of the ghost number of $A$ and of $B$ is equal to three, 
and the sum of the picture number of $A$ and of $B$ is equal to minus two:
\begin{equation} \label{relation:small}
\llangle A,B\rrangle = 0\qquad \text{unless}\qquad
\boldsymbol{g}(A) + \boldsymbol{g}(B) = 3\,,\quad
\boldsymbol{p}(A) + \boldsymbol{p}(B) = -2\,.
\end{equation}
Here $\boldsymbol{g}(A)$ and $\boldsymbol{g}(B)$ are the ghost number of $A$ and of $B$, respectively;
$\boldsymbol{p}(A)$ and $\boldsymbol{p}(B)$ are the picture number of $A$ and of $B$, respectively.
By contrast, the inner product in the large Hilbert space, $\langle A, B\rangle$,
vanishes unless the sum of the ghost numbers is equal to two
and the sum of the picture numbers is equal to minus one:
\begin{equation}
\langle A,B\rangle = 0\qquad \text{unless}\qquad
\boldsymbol{g}(A) + \boldsymbol{g}(B) = 2\,,\quad
\boldsymbol{p}(A) + \boldsymbol{p}(B) = -1\,.
\end{equation}
The two inner products are related as\footnote{
See appendix B of ref.~\cite{Torii:validity} for the reason why the imaginary unit is necessary.
}
\begin{equation} \label{small/large:BPZ}
\llangle A,B\rrangle = \iu\hs\langle\xz A,\hs B\rangle 
= \iu\hs (-1)^A \hs\langle A,\hs \xz B\rangle\qquad (\forall A, \forall B\in \Hs)\,,
\end{equation}
with
\begin{equation}
(-1)^A := \left\{
\begin{aligned}
&+1\quad \text{for $A$ bosonic}\\
&-1\quad \text{for $A$ fermionic}
\end{aligned}
\right.\,.
\end{equation}
Note that the inner product $\langle A,B\rangle$ in the large Hilbert space is identically zero
if both $A$ and $B$ are in the small Hilbert space:
\begin{equation} \label{prod=0}
\langle A,B\rangle = 0 \qquad (\forall A, \forall B\in \Hs)\,.
\end{equation}

%----------------------------------------
\subsection{The Witten formulation}
\label{Witten formulation}
The first formulation of manifestly covariant open superstring field theory was proposed by Witten~\cite{Witten:super}, 
based on the small Hilbert space approach.
It is a natural extension of the cubic open bosonic string field theory~\cite{Witten:bosonic},
with the action composed of the string fields in the \emph{natural} picture:
an NS string field of picture number minus one and a Ramond string field of picture number minus a half.
However, it has the problem of divergences caused by the picture-changing operator inserted at the string midpoint~\cite{Wendt}.
In the present subsection, we review this open superstring field theory and its problem, focusing on the NS sector.

The NS-sector action in the Witten formulation, $\SW$, is given by
\begin{equation} \label{Witten:NS-action}
\SW = -\frac{1}{2} \bllangle\PsiW , Q\PsiW \brrangle 
-\frac{g}{3} \bllangle \PsiW, \Xmid\bigl(\PsiW\ast\PsiW\bigr) \brrangle \,.
\end{equation}
Here $g$ denotes the open string coupling constant, 
$\Xmid$ denotes the picture-changing operator~\eqref{PCO} inserted at the string midpoint, 
$\PsiW$ is a Grassmann-odd NS open superstring field of even parity under the Gliozzi-Scherk-Olive (GSO) projection,
and the symbol ``$\ast$'' represents the multiplication in the space of string fields~\cite{Witten:bosonic}.
(All the superstring fields to appear in the present paper are GSO even.)
For later convenience, we have appended the superscript ``W'' to the string field in the Witten formulation.
The world-sheet ghost number $\boldsymbol{g}$ and the picture number $\boldsymbol{p}$ of $\PsiW$ are $+1$ and $-1$, respectively.
In what follows, the quantum number $(\boldsymbol{g},\boldsymbol{p})$ of a string field will often be indicated by its subscript. 
For example, the $\PsiW$ will be written also as $\PsiW_{(1,-1)}$.

As is mentioned in subsection~\ref{subsec:small/large}, 
inner products of the form $\llangle A,B\rrangle$ vanish unless $\boldsymbol{p}(A) + \boldsymbol{p}(B) = -2$.
Therefore, without the insertion of $X$, which raises picture number by one, the cubic term in the action \eqref{Witten:NS-action} 
would identically be zero, and the interacting theory could not be described.
On the other hand, the very midpoint insertion of $X$ causes the two serious problems:
scattering amplitudes are divergent even at the tree level, 
and gauge transformation is not well-defined~\cite{Wendt}.
Here we focus on the latter problem, which is related to the main subject of the present paper.
The gauge transformation in the Witten formulation is given by
\begin{equation} \label{singular gauge transf} 
\delta \PsiW_{(1,-1)} = Q \Lambda^\mrm{W}_{(0,-1)} 
+ g \Xmid \Bigl( \PsiW_{(1,-1)} \ast \Lambda^\mrm{W}_{(0,-1)} -  \Lambda^\mrm{W}_{(0,-1)} \ast \PsiW_{(1,-1)} \Bigr) \,,
\end{equation}
where $\Lambda^\mrm{W}_{(0,-1)}$ is a Grassmann-even gauge parameter of ghost number zero and picture number minus one.
(Note that in the Witten formulation, all the NS string fields are in the $-1$ picture.)
In the variation of the action under the above transformation,
the terms of order $g^0$ or $g^1$ vanish, and that of order $g^2$ takes the form
\begin{equation}
-g^2\hs \Bllangle \Xmid \Xmid \hs \bigl(\PsiW \ast \PsiW\bigr)\,,\, \PsiW\ast\LW - \LW\ast\PsiW \Brrangle\,.
\end{equation}
This would be zero if $\Xmid\Xmid$ were finite, but the fact is that the OPE of the picture-changing operator with itself is singular:
\begin{align} \label{XX}
X(z_1)X(z_2) &= \bigl\{Q, \xi(z_1)\bigr\} X(z_2) = \bigl\{Q, \xi(z_1)X(z_2)\bigr\} \nonumber \\*
&\sim
-\frac{2}{(z_1-z_2)^2}\bigl\{ Q, b\e^{2\phi}(z_2)\bigr\}
-\frac{1}{z_1 - z_2} \bigl\{ Q, \p (b\e^{2\phi})(z_2)\bigr\}\,.
\end{align}
Thus the product $\Xmid\Xmid$, in which two $X$'s collide at the string midpoint, is divergent,
and the gauge transformation \eqref{singular gauge transf} is not well-defined.

%----------------------------------------
\subsection{The Berkovits formulation}
\label{Berkovits formulation}
In order to remedy the problems in the Witten formulation, Berkovits has formulated open superstring field theory
without using any picture-changing operators~\cite{Berkovits:NS}.
This theory, unlike the Witten one, is constructed in the large Hilbert space.
The NS-sector action in the Berkovits formulation, $\SB$, takes the following Wess-Zumino-Witten form:\footnote{
A factor of the imaginary unit in each term is necessary in order for the action to be real. See appendix B of ref.~\cite{Torii:validity}.
}
\begin{equation} \label{S}
\SB = \frac{\iu}{2 g^2} \Bigl\langle
\Ginv \bigl( QG\bigr) \hs,\hs \Ginv \bigl( \ez G\bigr)\Bigr\rangle 
- \frac{\iu}{2 g^2} \int^1_0 \diff t\,\Bigl\langle
\bigl( \hatGinv \p_t \hatG \bigr) \hs,\hs 
\Bigl\{ \hatGinv\bigl( Q\hatG\bigr), \hatGinv \bigl(\ez \hatG\bigr)\Bigr\} \Bigr\rangle
\end{equation}
with 
\begin{equation} \label{G}
G = \exp \bigl( g\hs\Phi_{(0,0)}\bigr)\,,\quad
\hatG = \exp\bigl( t\hs g\hs\Phi_{(0,0)}\bigr)\,.
\end{equation}
Here $\Phi_{(0,0)}$ is a GSO-even NS string field whose Grassmann parity is even.
It carries no ghost number and no picture number, as is indicated by the subscript $(0,0)$.
In the above equations and in what follows, we omit the multiplication symbol ``$\ast$'' for simplicity,
but products of string fields are always defined by Witten's star product.

The operator $\ez$, as well as the BRST operator $Q$, acts as the derivation upon string fields, satisfying
\begin{equation} \label{Qeta} 
Q^2 = \ez^2 = \{ Q,\ez\} = 0\,.
\end{equation}
In virtue of this, the action \eqref{S} is invariant under the transformation of the form~\cite{tachyon}
\begin{equation} \label{gauge transf}
\delta G = g\Bigl[ \bigl( Q\hs\epsilon_{(-1,0)} \bigr) G + G\bigl( \ez\hs\epsilon_{(-1,1)}\bigr)\Bigr]\,.
\end{equation}
Note that there are not one but two gauge parameters $\epsilon_{(-1,0)}$ and $\epsilon_{(-1,1)}$.
As the result of the extension of the superstring Hilbert space, we have larger gauge symmetry in the Berkovits formulation than
in the Witten one.
Furthermore, the two parameters are in different pictures from each other.
In the Witten formulation, all the NS string fields are in the same picture,
and the picture-changing operation is realized by $X$,
whereas in the Berkovits formulation, picture numbers of string fields are not fixed to the same value.
Nevertheless, as shown in ref.~\cite{INOT}, the two formulations are related to each other:
if we perform partial gauge fixing in the Berkovits formulation,
the resultant action and the residual gauge transformation can be regarded as 
the regularized version of the action and of the gauge transformation in the Witten formulation.

%%%%%%%%%%%%%%%%%%%%%%%%%%%%%%%%%%%%%%%%%%%%%%%%%%%%%%
\section{Partial gauge fixing in the Berkovits formulation}
\label{section:Gauge-fixing}
\setcounter{equation}{0}
The two formulations of open superstring field theory introduced in the preceding section may look completely different.
They are, however, related to each other through partial gauge fixing~\cite{INOT}.
By fixing part of the gauge in the Berkovits formulation,
we can show that the free theories are equivalent; moreover, in the interacting case, 
the Berkovits formulation can be interpreted as the regularized version of the Witten one.
In the present section, we review this relation.
We first explain the basic idea of partial gauge fixing and demonstrate the equivalence of the two formulations
for the case of free theory in subsection~\ref{subsec:idea}.
Then we introduce in subsection~\ref{subsec:gfc} a one-parameter family of conditions for partial gauge fixing,
which is useful for the analysis of interacting theory.
After that, in subsection~\ref{residual symm}, we investigate the residual gauge symmetry under partial gauge fixing
and explain its relation to the Witten gauge transformation \eqref{singular gauge transf} 
in more detail and in a more sophisticated manner than in ref.~\cite{INOT}.

%----------------------------------------
\subsection{The basic idea of partial gauge fixing}
\label{subsec:idea}
Let us begin by reviewing the basic idea of partial gauge fixing.
For this purpose, we consider the free theories, showing their equivalence.
The equation of motion in the free Witten theory is given by
\begin{equation} \label{freeEOM:Witten}
Q\PsiW_{(1,-1)} = 0\,,
\end{equation}
and that in the free Berkovits theory is given by
\begin{equation} \label{freeEOM:Berkovits}
Q\ez\Phi_{(0,0)} = 0\,.
\end{equation}
Because the string field $\PsiW_{(1,-1)}$ in \eqref{freeEOM:Witten} is in the small Hilbert space, it satisfies
\begin{equation} \label{ezPsiW}
\ez\PsiW_{(1,-1)}= 0\,.
\end{equation}
Using this equation and the identity 
\begin{equation}
\{\xz , \ez \} = 1\,,
\end{equation}
we can rewrite \eqref{freeEOM:Witten} as
\begin{equation}
0 = Q\PsiW_{(1,-1)} = Q\{\xz, \ez\}\PsiW_{(1,-1)} = Q\ez\bigl(\xz\PsiW_{(1,-1)}\bigr) \,.
\end{equation}
Therefore, for any solution $\PsiW_{(-1,1)}$ to the equation of motion \eqref{freeEOM:Witten} in the Witten formulation,
we have a solution $\xz\PsiW_{(-1,1)}$ to the equation \eqref{freeEOM:Berkovits} in the Berkovits formulation.
In fact, by the use of the gauge transformation, 
the string field $\Phi_{(0,0)}$ in the Berkovits formulation can always be brought to the form
\begin{equation} \label{gauge-choice}
\Phi_{(0,0)} = \xz\Psi_{(1,-1)} 
\quad \text{with}\quad
\ez\Psi_{(1,-1)} = 0\,,
\end{equation}
where $\Psi_{(1,-1)}$ is some string field depending on $\Phi_{(0,0)}$.
Indeed, in the free theory,\footnote{For the case of interacting theory, see appendix~\ref{properness:pgf}.} 
the gauge transformation \eqref{gauge transf} reduces to
\begin{equation}
\delta\Phi_{(0,0)} = Q\epsilon_{(-1,0)} + \ez\epsilon_{(-1,1)}\,,
\end{equation}
and therefore if we consider the transformation specified by
\begin{equation}
\epsilon_{(-1,0)} = 0\,,\quad \epsilon_{(-1,1)} = -\xz\Phi_{(0,0)}\,,
\end{equation}
the resultant gauge transform takes the form
\begin{equation}
\Phi_{(0,0)} +\delta\Phi_{(0,0)} 
= \bigl(1-\ez\xz\bigr)\Phi_{(0,0)} = \xz\ez\Phi_{(0,0)}
=\xz\Psi_{(1,-1)}\,,
\end{equation}
with
\begin{equation}
\Psi_{(1,-1)} = \ez\Phi_{(0,0)}\,.
\end{equation}
In this manner, setting $\Phi_{(0,0)}$ in the form \eqref{gauge-choice} corresponds to fixing part of the gauge.
The condition for this partial gauge fixing is given by
\begin{equation} \label{xzPhi=0}
\xz\Phi_{(0,0)} = 0\,.
\end{equation}
Under this condition, we can show also that in free theory  the action in the Berkovits formulation
reduces to the gauge-invariant action in the Witten formulation.
To see this, let us start with the free Berkovits action
\begin{equation} \label{B:free}
\SB\bigl|_{g=0}\; = \frac{\iu}{2}\, \bigl\langle Q\Phi_{(0,0)}, \ez\Phi_{(0,0)}\bigr\rangle
= -\frac{\iu}{2}\, \bigl\langle \Phi_{(0,0)}, Q\ez\Phi_{(0,0)}\bigr\rangle\,.
\end{equation}
When $\Phi_{(0,0)}$ is written in the form \eqref{gauge-choice}, we obtain
\begin{equation}
-\frac{\iu}{2}\, \bigl\langle \Phi_{(0,0)}, Q\ez\Phi_{(0,0)}\bigr\rangle
= -\frac{\iu}{2}\, \bigl\langle \xz\Psi_{(1,-1)}, Q\Psi_{(1,-1)}\bigr\rangle
= -\frac{1}{2} \,\bllangle \Psi_{(1,-1)}, Q\Psi_{(1,-1)}\brrangle\,.
\end{equation}
(In the last equality, we have used the relation \eqref{small/large:BPZ}.)
Thus \eqref{B:free} coincides with the free Witten action under the identification
\begin{equation}
\Psi_{(1,-1)} \cong \PsiW_{(1,-1)}\,.
\end{equation}

In the above argument, we have only used the following properties of $\xz$:
\begin{equation}
(\boldsymbol{g},\boldsymbol{p}) = (-1,1)\,,\quad
\xz^2 = 0\,,\quad
\{\xz, \ez\} = 1\,.
\end{equation}
Therefore we may replace $\xz$ with a Grassmann-odd operator $\Xi$ satisfying
\begin{equation}
(\boldsymbol{g},\boldsymbol{p}) = (-1,1)\,,\quad
\Xi^2 = 0\,,\quad
\{\Xi, \ez\} = 1\,,
\end{equation}
and may consider the condition 
\begin{equation} \label{XiPhi=0}
\Xi\,\Phi_{(0,0)} = 0\,.
\end{equation}
The relation \eqref{small/large:BPZ} is then generalized to
\begin{equation} \label{small/large:BPZ:Xi}
\llangle A,B\rrangle = \iu\hs\langle\Xi A,\hs B\rangle \,,\quad
\llangle A,B\rrangle = \iu\hs (-1)^A \hs\langle A ,\hs \Xi B\rangle\qquad (\forall A, \forall B\in \Hs)\,,
\end{equation}
and we can show the equivalence of the two free theories under the partial gauge fixing \eqref{XiPhi=0} 
in the same manner as before,
identifying $\Psi_{(1,-1)} = \ez\Phi_{(0,0)}$ in the Berkovits formulation 
with $\PsiW_{(1,-1)}$ in the Witten formulation.

%----------------------------------------------------------------------
\subsection{A one-parameter family of conditions for partial gauge fixing}
\label{subsec:gfc}
In the preceding subsection, we examined only the free theories.
To show their equivalence, we did not have to specify the form of $\Xi$ in \eqref{XiPhi=0}.
In the interacting case, however, the choice of $\Xi$ becomes important.
A particular type of $\Xi$ helps us to manifest the relation between the two formulations.
In the present subsection, we review such useful gauge choices proposed in ref.~\cite{INOT}.

We consider a one-parameter family of conditions for partial gauge fixing of the form
\begin{equation} \label{gfc}
\Xil\Phi_{(0,0)} =0\qquad (0<\lambda <\infty)\,.
\end{equation}
Here $\Xil$ are operators defined by integrals along the counterclockwise unit circle $C$ centered at the origin:
\begin{equation} \label{Xi,u}
\Xil = \oint_C \frac{\diff z}{2\pi\iu} u_\lambda (z) \xi(z)\,,
\end{equation}
with
\begin{equation} \label{u:even}
u_\lambda (z) = \frac{1}{z-\iu\e^{-\lambda}} - \frac{1}{z-\iu\e^{\lambda}}\,.
\end{equation} 
As is explained in ref.~\cite{INOT}, in the limit $\lambda\to 0$ we have
\begin{equation} \label{Xil}
\Xil = \xi(\iu \e^{-\lambda}) + O(\lambda)\,.
\end{equation}
From the viewpoint of the state-operator correspondence in the conformal frame on the upper half-plane,
the open string lies on the unit upper half-circle centered at the origin,
with its midpoint at $z=\iu$.
Therefore, the operator $\Xil$ approaches $\xi_\mathrm{mid}$, the midpoint insertion of $\xi$, as the parameter $\lambda$ tends to zero:
\begin{equation}
\Xil \to \xi_\mrm{mid} \quad (\lambda \to 0)\,.
\end{equation}
Furthermore, from \eqref{Xil} we obtain
\begin{equation}
\Xl = X(\iu \e^{-\lambda}) + O(\lambda)\,,
\end{equation}
where $\Xl$ is the BRST transform of $\Xil$:
\begin{equation} \label{calX}
\Xl := \{ Q, \Xil\}\,.
\end{equation}
Hence $\Xl$ becomes $X_\mrm{mid}$, the midpoint insertion of the picture-changing operator, when $\lambda$ goes to zero:
\begin{equation}
\Xl \to X_\mrm{mid} \quad (\lambda \to 0)\,.
\end{equation}
We also note that $\Xil$ are BPZ even:
\begin{equation} \label{BPZ:Xil}
\mrm{bpz}\bigl(\Xil\bigr) = \Xil\,.
\end{equation}

%------------------------------------------------------
\subsection{Residual gauge symmetry under partial gauge fixing}
\label{residual symm}
The condition \eqref{gfc} considered in the preceding subsection 
eliminates the gauge degrees of freedom in the Berkovits formulation only partially.
Therefore, there remains residual gauge symmetry even after the condition is imposed.
In the present subsection, we investigate the residual gauge transformation which preserves condition \eqref{gfc},
which can be regarded as the regularized version of the Witten gauge transformation \eqref{singular gauge transf},
in more detail and in a more sophisticated manner than in ref.~\cite{INOT}.
For this purpose, 
it is convenient to express \eqref{gauge transf} in terms of $\Phi = \Phi_{(0,0)}$ rather than $G$.
In order to perform this rewriting, we introduce the adjoint operator $\adg = g\,\adPhi$,
whose action upon a string field $A$ is defined by
\begin{equation}
\adg (A) \equiv g\,\adPhi (A) := g\left[\Phi, A\right] .
\end{equation}
This operator satisfies
\begin{equation}
\e^{\alpha\hs \adg} A = \exp\bigl(\alpha\hs g\hs \Phi\bigr) A \exp\bigl(-\alpha\hs g\hs \Phi\bigr)
\qquad \left(\forall \alpha \in \mathbb{C}\right)\,.
\end{equation}
From $\left[ \Phi, G \right] = 0$, for an arbitrary variation of $\Phi$ we obtain
\begin{align} \label{delta:Phi}
\e^{-\frac{1}{2}g\Phi} \delta\left[\Phi, G\right] \e^{-\frac{1}{2}g\Phi} = 0 
&\iff \bigl( \e^{\frac{1}{2}\adg} -\e^{-\frac{1}{2}\adg}\bigr)\delta\Phi = \e^{-\frac{1}{2}g\Phi} \left[\Phi, \delta G\right] \e^{-\frac{1}{2}g\Phi} 
\nonumber \\
&\iff \delta\Phi = g^{-1}\frac{\adg}{1-\e^{-\adg}} \bigl( \Ginv \delta G\bigr) .
\end{align}
Therefore, the gauge variation of $\Phi$ is given by
\begin{align} \label{dP} 
\delta \Phi
&= \frac{\adg}{1-\e^{-\adg}}\, \Bigl( \e^{-\adg} Q\hs\epsilon_{(-1,0)} + \ez\hs\epsilon_{(-1,1)} \Bigr) 
= \frac{\adg}{\e^{\adg}-1}\,Q\epsilon_{(-1,0)} + \frac{-\adg}{\e^{-\adg}-1}\, \ez\hs\epsilon_{(-1,1)} \,.
\end{align}
Now suppose that $\Phi$ obeys condition \eqref{gfc}.
In order for the gauge transform of $\Phi$ to keep the condition,
the variation \eqref{dP} has to satisfy
\begin{equation} \label{Xil:delta}
\Xil \, \delta \Phi = 0\,.
\end{equation}
Under this constraint, the gauge parameters $\epsilon_{(-1,0)}$ and $\epsilon_{(-1,1)}$ are not independent any longer.
As a matter of fact, we can express $\ez\epsilon_{(-1,1)}$ in terms of $\Phi$ and $Q\epsilon_{(-1,0)}$ as follows.
Because the residual transformation satisfies \eqref{Xil:delta}, we have
\begin{equation}
\ez\epsilon_{(-1,1)} = \ez\Xil\Bigl(\ez\epsilon_{(-1,1)} - \delta \Phi\Bigr) 
= -\ez\Xil\biggl[ \frac{\adg}{\e^{\adg} -1}Q\epsilon_{(-1,0)} +\biggl(\frac{-\adg}{\e^{-\adg}-1} -1\biggr)\ez\epsilon_{(-1,1)}\biggr].
\end{equation}
This equation can be solved recursively in $\ez\epsilon_{(-1,1)}$, and we obtain
\begin{align} \label{recursion}
\ez\epsilon_{(-1,1)} &= -\sum^\infty_{n=0}\biggl[ -\ez\Xil\biggl(\frac{-\adg}{\e^{-\adg}-1} -1\biggr)\biggr]^n
\ez\Xil\,\frac{\adg}{\e^{\adg}-1}\,Q\epsilon_{(-1,0)}
\nonumber \\[.5ex]
&= - \biggl[ 1 + \ez\Xil\biggl(\frac{-\adg}{\e^{-\adg}-1} -1\biggr)\biggr]^{-1}
\ez\Xil\,\frac{\adg}{\e^{\adg}-1}\,Q\epsilon_{(-1,0)}\,. 
\end{align}
In the last equality, we have used the fact that 
the sum in \eqref{recursion} converges for small $g$
because the operator 
\begin{equation}
-\ez\Xil\biggl(\frac{-\adg}{\e^{-\adg}-1} -1\biggr) 
\end{equation}
is $O(g)$.
Substituting \eqref{recursion} into \eqref{dP},
we find that the explicit form of the residual transformation is given by
\begin{equation} \label{explicit}
\delta_\mrm{res} \Phi 
= \Biggl( 1 - \frac{-\adg}{\e^{-\adg}-1} \biggl[ 1 + \ez\Xil\biggl(\frac{-\adg}{\e^{-\adg}-1} -1\biggr)\biggr]^{-1} \ez\Xil
\Biggr)  \frac{\adg}{\e^{\adg}-1}\,Q\epsilon_{(-1,0)}\,.
\end{equation}
Thus the residual gauge transformation in the Berkovits formulation is parameterized by a single gauge parameter
as the gauge transformation in the Witten formulation is.
In fact, the former can be regarded as the regularized version of the latter.
To see this, we express $\Phi$ as
\begin{equation} \label{Phi:Xil:Psi}
\Phi = \Xil \Psi\,,\quad \Psi \in \Hs\,,
\end{equation}
and consider the residual transformation in terms of the string field $\Psi = \ez\Phi$,
which will correspond to $\PsiW_{(1,-1)}$ in the Witten formulation.
Expanding \eqref{explicit} in $g$, we obtain
\begin{equation} \label{Psi:transf}
\delta_\mrm{res} \Psi 
= -Q \ez\epsilon_{(-1,0)}
 - \frac{g}{2}\Bigl( \bigl[ \Psi\,,\, (1 +\ez\Xil)\, Q\epsilon_{(-1,0)} \bigr] - \bigl[ \Xil\Psi\,,\, Q \ez\epsilon_{(-1,0)} \bigr] \Bigr)
+ O(g^2)\,.
\end{equation}
Unlike the gauge parameter $\Lambda^\mrm{W}_{(0,-1)}$ in the Witten formulation, which is in the small Hilbert space,
the parameter $\epsilon_{(-1,0)}$ in the Berkovits formulation moves around the whole of the large Hilbert space.
Nevertheless, as shown in appendix~\ref{properness}, without loss of generality we may assume that $\epsilon_{(-1,0)}$
satisfies the condition
\begin{equation} \label{Xil:LQ}
\Xil\hs\epsilon_{(-1,0)} =0\,.
\end{equation}
Therefore, $\epsilon_{(-1,0)}$ can be written as
\begin{equation} \label{LambdaQ}
\epsilon_{(-1,0)} = -\Xil\Lambda_{(0,-1)} 
\end{equation}
for some $\Lambda_{(0,-1)}\in\Hs$,
and the $\Lambda_{(0,-1)}$ is naturally related to the gauge parameter in the Witten formulation.
This can be seen as follows.
First, we transform the term $(1+\ez\Xil) Q\epsilon_{(-1,0)}$ in \eqref{Psi:transf} as
\begin{align}
\left(1+\ez\Xil\right) Q \epsilon_{(-1,0)}
& = -Q \Xil \Lambda_{(0,-1)} -\ez \Xil Q \Xil \Lambda_{(0,-1)}
= -Q \Xil \Lambda_{(0,-1)} -\ez \Xl \Xil \Lambda_{(-1,0)} \nonumber \\[.5ex]
&= -Q \Xil \Lambda_{(0,-1)} -\Xl \ez \Xil \Lambda_{(0,-1)} 
= -Q \Xil \Lambda_{(0,-1)} -\Xl \Lambda_{(0,-1)} \nonumber \\[.5ex]
&= -2 \Xl \Lambda_{(0,-1)} +\Xil Q \Lambda_{(0,-1)} \,.
\end{align}
Here we have used the relations
\bs
\begin{align}
\{ Q, \Xil \} &= \Xl\,,\\[.5ex]
\Xil^2 &= 0\,,\\[.5ex]
\left[ \ez, \Xl \right] &= 0\,, \label{ezXl=0} \\[.5ex]
\ez \Xil \Lambda_{(0,-1)} &= \left( 1 -\Xil \ez \right) \Lambda_{(0,-1)} = \Lambda_{(0,-1)}\,.
\end{align}
\es
Then, eq.~\eqref{Psi:transf} becomes
\begin{equation} \label{var}
\delta_\mrm{res} \Psi 
= Q\Lambda_{(0,-1)}
- \frac{g}{2}\Bigl( \bigl[ \Psi\,,\, (-2\Xl + \Xil Q)\Lambda_{(0,-1)} \bigr]
+ \bigl[ \Xil\Psi\,,\, Q\Lambda_{(0,-1)} \bigr] \Bigr) 
+ O(g^2)\,.
\end{equation}
The point is that in the singular limit $\lambda\to 0$, 
the operators $\Xl$ and $\Xil$ in \eqref{var} effectively become the midpoint insertions $X_\mrm{mid}$ and $\xi_\mrm{mid}$, respectively,
and we have
\begin{align} \label{limit:res}
\delta_\mrm{res}\Psi 
&\overset{\lambda\to 0}{\longrightarrow} 
Q\Lambda_{(0,-1)} + g\Bigl( \bigl[\Psi, X_\mrm{mid} \Lambda_{(0,-1)}\bigr] 
- \bigl[\Psi, \xi_\mrm{mid} Q\Lambda_{(0,-1)}\bigr] - \bigl[\xi_\mrm{mid} \Psi,Q\Lambda_{(0,-1)}\bigr] \Bigr)
+O(g^2) \nonumber \\[.5ex]
&= Q\Lambda_{(0,-1)} + g X_\mrm{mid}\bigl[\Psi,\Lambda_{(0,-1)}\bigr] + O(g^2)\,.
\end{align} 
In the last line, we have used 
\begin{align}
\bigl( X_\mrm{mid} A\bigr) B &= A \bigl(X_\mrm{mid} B\bigr) = X_\mrm{mid} \bigl( AB\bigr)\,, \\[1ex]
\bigl( \xi_\mrm{mid} A\bigr) B &= (-1)^A A \bigl(\xi_\mrm{mid} B\bigr) = \xi_\mrm{mid} \bigl( AB\bigr)
\end{align}
for any pair of string fields $A$ and $B$.
It should be noted that the $O(g^2)$ term in \eqref{limit:res} is divergent.
Apart from this $O(g^2)$ term, however, the last line of \eqref{limit:res}
precisely coincides with the Witten gauge transformation \eqref{singular gauge transf} under the identification
\begin{equation}
\Psi\cong \PsiW_{(1,-1)}\,,\quad
\Lambda_{(0,-1)}\cong \Lambda^\mrm{W}_{(0,-1)}\,.
\end{equation}
Thus we find that in this singular limit, the residual transformation \eqref{Psi:transf}
is naturally related to the Witten gauge transformation.
We could say that the $O(g^2)$ term plays the role of the counterterm.
In fact, the Berkovits theory under the gauge \eqref{gfc} can be regarded as the regularized version of the Witten one~\cite{INOT}.

%%%%%%%%%%%%%%%%%%%%%%%%%%%%%%%%%%%%%%%%%%%%%%%%%%%%%%%
\section{Reducibility structure}
\label{sec:reducibility}
\setcounter{equation}{0}
In subsection~\ref{residual symm}, we have considered the residual gauge transformation under partial gauge fixing of the Berkovits theory,
manifesting its relation to the Witten gauge transformation.
Developing our analysis further, in the present section, we investigate more detailed gauge structure called reducibility structure.
The structure is important in that it governs quantization procedure.
For example, it determines whether the quantization requires not only ordinary ghosts but also additional ghosts such as ghosts for ghosts.

We first explain the concept of reducibility in subsection~\ref{concept},
and illustrate reducibility structure with open bosonic string field theory in subsection~\ref{red:bosonic}.
We then review the reducibility structure of the Witten formulation in subsection~\ref{red:Witten},
and explain that of the Berkovits formulation in subsection~\ref{red:Berkovits}.
After that, in subsection~\ref{relation:reducibility}, we examine the relation between the reducibility structures
of the two formulations.
We find that the reducibility structure of the Berkovits formulation under the partial gauge fixing includes, 
as a sub-structure, the regularized version of the reducibility structure of the Witten formulation. 

%--------------------------------------------------------------------------
\subsection{The concept of reducibility}
\label{concept} 
In order to explain the concept of reducibility,
let us consider a gauge system described by a classical action $S = S[\varphi]$, 
which depends on a classical field $\varphi$,
and assume that the action is invariant under gauge transformation of the form\footnote{
For simplicity, we consider a gauge system described by a single field $\varphi$ and a single gauge parameter $\epsilon_0$.
}
\begin{equation}
\varphi \ \longrightarrow\ \varphi + \delta_{\epsilon_0} \varphi \,.
\end{equation}
In the above equation, we have appended the subscript ``$\epsilon_0$'' to the variation symbol $\delta$ 
in order to indicate that the variation is parameterized by $\epsilon_0$.
We use a similar notation in what follows in the present subsection (and only in the present subsection).
In general, if we change the parameter $\epsilon_0$ as 
\begin{equation} \label{shift}
\epsilon_0\ \longrightarrow\ \epsilon_0 + \delta_{\epsilon_1}\epsilon_0\,,
\end{equation}
with $\epsilon_1$ parameterizing the variation of $\epsilon_0$,
the form of $\delta_{\epsilon_0} \varphi$ changes accordingly.
In some theories, however, for some particular choice of $\delta_{\epsilon_1}\epsilon_0$,
the gauge variation $\delta_{\epsilon_0} \varphi$ is invariant under \eqref{shift} \emph{if we use the equation of motion}
\begin{equation} \label{red:eom}
\frac{\delta S}{\delta\varphi} = 0\,.
\end{equation}
For such $\delta_{\epsilon_1}\epsilon_0$, we have
\begin{equation} \label{d1d0}
\delta_{\epsilon_1} \bigl( \delta_{\epsilon_0}\varphi\bigr) \simeq 0\,,
\end{equation} 
where the symbol ``$\simeq$'' denotes the equality which holds if (but not necessarily only if) 
we use the equation of motion.
Furthermore, in complicated gauge systems,
there exists also a variation
\begin{equation}
\epsilon_1 \ \longrightarrow\ \epsilon_1 + \delta_{\epsilon_2}\epsilon_1
\end{equation} 
which keeps $\delta_{\epsilon_1}\epsilon_0$ invariant
under the use of the equation of motion:
\begin{equation}
\delta_{\epsilon_2}\bigl(\delta_{\epsilon_1}\epsilon_0\bigr) \simeq 0\,.
\end{equation}
In general, a gauge system is said to be $N$-th stage reducible if there exist a series of parameters $\epsilon_k$ ($0\leq k \leq N$)
and a series of variations $\delta_{\epsilon_k}\epsilon_{k-1}$ ($0\leq k\leq N\hs ; \; \epsilon_{-1} := \varphi$) which satisfy
\begin{equation}
\delta_{\epsilon_{k}}\bigl( \delta_{\epsilon_{k-1}} \epsilon_{k-2}\bigr) \simeq 0
\quad \left(1\leq k\leq N\right).
\end{equation}
As we see in the following subsections, string field theory is infinitely reducible,
with the above $N$ infinity.

%-----------------------------------------------------------------------------
\subsection{Reducibility structure of open bosonic string field theory}
\label{red:bosonic}
To illustrate reducibility structure, let us consider open bosonic string field theory~\cite{Witten:bosonic}, for example.
Its action takes the form
\begin{equation} \label{bosonic-action:free}
S_\mrm{bos} = -\frac{1}{2} \langle\Psi_1 , Q\Psi_1 \rangle 
-\frac{g}{3} \langle \Psi_1, \Psi_1 \ast \Psi_1 \rangle \,,
\end{equation}
where $\Psi_1$ is a Grassmann-odd string field of world-sheet ghost number one.
In what follows, the world-sheet ghost number $\boldsymbol{g}$ of a bosonic string field is indicated by a subscript 
as in $\Psi_{\boldsymbol{g}}$. 
The BRST operator and the BPZ inner product in the Hilbert space of the bosonic theory are denoted by
the same symbols $Q$ and $\langle\ \,,\,\ \rangle$, respectively, as in the Berkovits formulation, for simplicity.

Let us first examine the free theory.
The gauge transformation is given by
\begin{equation} \label{del0:bosonic}
\delta_0 \Psi_1 = Q\Lambda_0\,.
\end{equation}
For convenience, we have appended the subscript ``$0$'' to the variation symbol $\delta$.
The point is that because $Q$ is nilpotent, $\delta_0 \Psi_1$ does not change under the following variation of the gauge parameter:
\begin{equation} \label{del1:bosonic}
\delta_1 \Lambda_0 = Q\Lambda_{-1}\,.
\end{equation}
Moreover, eq.~\eqref{del1:bosonic} is invariant under the variation of the form
\begin{equation} \label{del2:bosonic}
\delta_2 \Lambda_{-1} = Q\Lambda_{-2}\,.
\end{equation}
In fact, there exists a series of variations of parameters.
The $n$-th variation
\begin{equation} \label{deln:bosonic}
\delta_n \Lambda_{-(n-1)} = Q\Lambda_{-n}
\qquad \bigl(n\geq 0\hs ;\,\ \Lambda_1 := \Psi_1\bigr)
\end{equation}
is not affected by the $(n+1)$-st variation of the form
\begin{equation}
\delta_{n+1} \Lambda_{-n} = Q\Lambda_{-(n+1)}\,.
\end{equation}
In other words, we have
\begin{equation}
\delta_{n+1}\Bigl( \delta_n \Lambda_{-(n-1)} \Bigr) = 0\qquad \left(n\geq 0\right).
\end{equation}
The Grassmann parity of $\Lambda_{-n}$ is even (resp.\ odd) if $n$ is even (resp.\ odd).
Note that the above equation holds without using the equation of motion.
This is a feature of the free string field theory.
In the interacting theory, however, this is not the case.
Let us next see this.

By the presence of the interaction, the gauge transformation \eqref{del0:bosonic}
is replaced with
\begin{equation} \label{del0:bosonic:full}
\delta_0 \Psi_1 = Q\Lambda_0 + g \bigl[ \Psi_1, \Lambda_0 \bigr] \,.
\end{equation}
Unlike the free theory, the interacting theory does not have a variation $\delta_1\Lambda_0$
which keeps $\delta_0\Psi_1$ strictly invariant. 
However, we find that the variation
\begin{equation} \label{del1:bosonic:full}
\delta_1\Lambda_0 = Q\Lambda_{-1} + g\bigl\{\Psi_1, \Lambda_{-1}\bigr\}
\end{equation}
does not change \eqref{del0:bosonic:full} if we use the equation of motion
\begin{equation} \label{EOM:bosonic}
\frac{\delta S_\mrm{bos}}{\delta\Psi_1} \equiv -Q\Psi_1 - g\bigl( \Psi_1\ast\Psi_1\bigr) = 0\,.
\end{equation}
Indeed we have
\begin{equation}
\delta_1\bigl(\delta_0\Psi_1\bigr)
= -g \left[\frac{\delta S_\mrm{bos}}{\delta\Psi_1}\,,\, \Lambda_{-1}\right],
\end{equation}
which vanishes under the use of the equation of motion.
Moreover, eq.~\eqref{del1:bosonic:full} is invariant under the variation
\begin{equation}
\label{del2:bosonic:full}
\delta_2\Lambda_{-1} = Q\Lambda_{-2} + g\bigl[\Psi_1, \Lambda_{-2}\bigr]
\end{equation}
if we use \eqref{EOM:bosonic}.
In fact, if the equation of motion holds, the $n$-th variation in the interacting theory
\begin{equation} \label{deln:bosonic:full}
\delta_n\Lambda_{-(n-1)} = Q\Lambda_{-n} + g \bigl[\Psi_1 , \Lambda_{-n}\bigr\}
\quad \left(n\geq 0\right)
\end{equation} 
with $[\,\ ,\,\, \}$ denoting the graded commutator
is not affected by the $(n+1)$-st variation of the form
\begin{equation}
\delta_{n+1} \Lambda_{-n} = Q\Lambda_{-(n+1)} + g \bigl[\Psi_1 , \Lambda_{-(n+1)}\bigr\}
\end{equation}
because we have
\begin{equation}
\delta_{n+1}\Bigl(\delta_n\Lambda_{-(n-1)}\Bigr)
= -g \left[\frac{\delta S_\mrm{bos}}{\delta\Psi_1}\,,\, \Lambda_{-(n+1)}\right].
\end{equation}
We thus find that the open bosonic string field theory is infinitely reducible.
It should be noted that it is because $g$ is zero that in the free theory
$\delta_{n+1}\bigl(\delta_n\Lambda_{-(n-1)}\bigr)$ exactly vanish.

%-----------------------------------------------------------------------------
\subsection{Reducibility structure of the Witten formulation}
\label{red:Witten}
As we have seen in subsection~\ref{Witten formulation}, 
open superstring field theory in the Witten formulation has the problem of divergences caused by the collision of
picture-changing operators.
Apart from such divergences, however, its reducibility structure is quite similar to 
that of open bosonic string field theory.
In particular, the structures of the free theories are exactly the same.
Indeed, the gauge variation
\begin{equation} \label{del0:Witten}
\delta_0 \PsiW_{(1,-1)} = Q\LW_{(0,-1)}
\end{equation}
in the free open superstring field theory
does not change under the variation of the form
\begin{equation} \label{del1:Witten}
\delta_1\LW_{(0,-1)} = Q\LW_{(-1,-1)}\,,
\end{equation}
and the $n$-th variation
\begin{equation}
\delta_n \LW_{(-(n-1), -1)} = Q\LW_{(-n,-1)}
\qquad \bigl(n\geq 0\hs ;\,\ \LW_{(1,-1)} := \PsiW_{(1,-1)}\bigr)
\end{equation}
is not affected by the $(n+1)$-st variation 
\begin{equation}
\delta_{n+1} \LW_{(-n,-1)} = Q\LW_{(-(n+1),-1)}\,.
\end{equation}
Thus we have
\begin{equation}
\delta_{n+1}\Bigl( \delta_n \LW_{(-(n-1), -1)} \Bigr) = 0\qquad \left(n\geq 0\right),
\end{equation}
where the Grassmann parity of $\LW_{(-n,-1)}$ is even (resp.\ odd) if $n$ is even (resp.\ odd).

Let us next consider the interacting NS-sector theory.
In this case, there arise divergences caused by the picture-changing operator,
but we examine formal gauge structure, neglecting such divergences. 
In the interacting theory, the gauge transformation is given by
\begin{equation} \label{del0:Witten:full}
\delta_0 \PsiW_{(1,-1)} = Q\LW_{(0,-1)} + g\Xmid\Bigl[ \PsiW_{(1,-1)}, \LW_{(0,-1)}\Bigr] .
\end{equation}
As in the case of the bosonic theory, the interacting NS-sector theory does not have a variation $\delta_1\LW_{(0,-1)}$
which keeps $\delta_0\PsiW_{(1,-1)}$ strictly invariant. 
However, the variation 
\begin{equation} \label{del1:Witten:full}
\delta_1\LW_{(0,-1)} = Q\LW_{(-1,-1)} + g\Xmid\Bigl\{\PsiW_{(1,-1)}, \LW_{(-1,-1)}\Bigr\}
\end{equation}
does not change \eqref{del0:Witten:full} \emph{formally} if we use the equation of motion
\begin{equation} \label{EOM:Witten}
\frac{\delta\SW}{\delta\PsiW} \equiv -Q\PsiW - g\Xmid\bigl( \PsiW\ast\PsiW \bigr)= 0\,.
\end{equation}
Indeed we have
\begin{equation}
\delta_1\bigl(\delta_0\PsiW_{(1,-1)}\bigr)
= -g\Xmid\left[\frac{\delta\SW}{\delta\PsiW}\,,\, \LW_{(-1,-1)}\right],
\end{equation}
which vanishes under the use of the equation of motion if we neglect the divergence caused by
the $\Xmid$ in front of the commutator and that in $\frac{\delta\SW}{\delta\PsiW}$.
Similarly, if we use \eqref{EOM:Witten}, the $n$-th variation
\begin{equation} \label{deln:Witten:full}
\delta_n\LW_{(-(n-1),-1)} = Q\LW_{(-n,-1)} + g\Xmid\Bigl[\PsiW_{(1,-1)}, \LW_{(-n,-1)}\Bigr\}
\quad \left(n\geq 0\right)
\end{equation}
is not affected formally by the $(n+1)$-st variation
\begin{equation}
\delta_{n+1} \LW_{(-n,-1)} = Q\LW_{(-(n+1),-1)} + g\Xmid\Bigl[\PsiW_{(1,-1)}, \LW_{(-(n+1),-1)}\Bigr\}
\end{equation}
because we have
\begin{equation}
\delta_{n+1}\Bigl(\delta_n\LW_{(-(n-1),-1)}\Bigr)
= -g\Xmid\left[\frac{\delta\SW}{\delta\PsiW}\,,\, \LW_{(-(n+1),-1)}\right].
\end{equation}

%--------------------------------------------------------------------------
\subsection{Reducibility structure of the Berkovits formulation}
\label{red:Berkovits}
Now that we have examined the reducibility structure of the Witten formulation,
let us next consider that of the Berkovits formulation.
Here we will review the structure before partial gauge fixing, 
following refs.~\cite{paperI, paperII}.
As we will see, the Berkovits theory, also, is infinitely reducible.

Let us begin by analyzing the free theory. The gauge transformation in the free theory is given by
\begin{equation} \label{del0:Phi}
\delta_0 \Phi = Q\hs\epsilon_{(-1,0)} + \ez\hs\epsilon_{(-1,1)}\,.
\end{equation}
In the matrix notation, this can be written as
\begin{equation} \label{del0:Berkovits}
\delta_0 \Phi =
\bbm
Q&\ez
\ebm
\bbm
\epsilon_{(-1,0)} \\[1ex]
\epsilon_{(-1,1)}
\ebm
.
\end{equation}
In virtue of the relation
\begin{equation}
Q^2 = \ez^2 = \{Q, \ez\} =0 \,,
\end{equation}
eq.~\eqref{del0:Berkovits} does not change under the variation of the form
\begin{equation} \label{del1:Berkovits}
\delta_1
\bbm
\epsilon_{(-1,0)} \\[1ex]
\epsilon_{(-1,1)}
\ebm
= 
\bbm
Q&\ez &0 \\[.5ex]
0&Q&\ez
\ebm
\bbm
\epsilon_{(-2,0)} \\
\epsilon_{(-2,1)} \\
\epsilon_{(-2,2)}
\ebm
.
\end{equation}
Moreover, eq.~\eqref{del1:Berkovits} is invariant under the variation
\begin{equation}
\delta_2 
\bbm
\epsilon_{(-2,0)} \\
\epsilon_{(-2,1)} \\
\epsilon_{(-2,2)}
\ebm
=
\bbm
Q&\ez &0 &0\\[.5ex]
0&Q&\ez &0 \\[.5ex]
0&0&Q&\ez
\ebm
\bbm
\epsilon_{(-3,0)} \\
\epsilon_{(-3,1)} \\
\epsilon_{(-3,2)} \\
\epsilon_{(-3,3)}
\ebm
.
\end{equation}
In fact, there exists a series of variations of parameters.
The $n$-th variation
\begin{equation} \label{deln:Berkovits}
\delta_n
\begin{bmatrix}
\epsilon_{(-n, 0)} \\
\vdots \\
\epsilon_{(-n, n)}
\end{bmatrix}
= \ n+1 \underbrace{\left\{
\begin{bmatrix}
Q & \eta_0 & & \hsymbu{0} \\[.5ex]
& \ddots & \ddots & \\[.5ex]
\hsymbl{0} & & Q & \eta_0 \\
\end{bmatrix}
\right.}_{n+2}
\begin{bmatrix}
\epsilon_{(-(n+1), 0)} \\
\vdots \\
\epsilon_{(-(n+1), n+1)}
\end{bmatrix} 
\quad ( n\geq 0)
\end{equation}
with 
\begin{equation}
\epsilon_{(0,0)} := \Phi
\end{equation}
is not affected by the $(n+1)$-st variation of the form
\begin{equation} 
\delta_{n+1}
\begin{bmatrix}
\epsilon_{(-(n+1), 0)} \\
\vdots \\
\epsilon_{(-(n+1), n+1)}
\end{bmatrix}
= \ n+2 \underbrace{\left\{
\begin{bmatrix}
Q & \eta_0 & & \hsymbu{0} \\[.5ex]
& \ddots & \ddots & \\[.5ex]
\hsymbl{0} & & Q & \eta_0 
\end{bmatrix}
\right. }_{n+3}
\begin{bmatrix}
\epsilon_{(-(n+2), 0)} \\
\vdots \\
\epsilon_{(-(n+2), n+2)}
\end{bmatrix} 
\,.
\end{equation}
In other words, we have
\begin{equation}
\delta_{n+1}\left(\delta_n
\begin{bmatrix}
\epsilon_{(-n, 0)} \\
\vdots \\
\epsilon_{(-n, n)}
\end{bmatrix}
\right)
=0
\quad ( n\geq 0)\,.
\end{equation}
The Grassmann parity of $\epsilon_{(-n,m)}$ ($0\leq m\leq n$) is even (resp.\ odd) if $n$ is even (resp.\ odd).
As in the case of the free Witten theory, 
the above equation holds without the use of the equation of motion. 

Next let us consider the reducibility structure of the interacting theory.
For this purpose, it is convenient to introduce the deformed BRST operator \cite{paperII, Torii:proceedings}
\begin{equation}
\tQ := \e^{-\adg}Q\e^{\adg}\,.
\end{equation}
This operator is nilpotent as $Q$ is:
\begin{equation}
\tQ^2 = 0\,.
\end{equation}
However, it does not anticommute with $\ez$: for an arbitrary string field $A$, we have
\begin{equation} \label{tQez:A}
\bigl\{\tQ, \ez\bigr\} A = -\iu\, g^2\left[\frac{\delta \SB}{\delta G}G\hs ,\hs A\right]. 
\end{equation}
The right-hand side is proportional to the derivative of the action\footnote{
For the derivation of \eqref{derivative}, see ref.~\cite{tachyon} for example.
}
\begin{equation} \label{derivative}
\frac{\delta\SB}{\delta G} = \frac{\iu}{g^2}\, \ez\left(\Ginv QG\right)\Ginv\,,
\end{equation}
and therefore we obtain
\begin{equation} \label{tQez}
\bigl\{\tQ,\ez\bigr\} \simeq 0\,.
\end{equation}
In order to investigate the reducibility structure, 
we start from the following form of the gauge transformation (see \eqref{dP}):
\begin{equation} \label{delP}
\delta_0 \Phi
= \frac{\adg}{1-\e^{-\adg}}\, \Bigl( \e^{-\adg} Q\hs\epsilon_{(-1,0)} + \ez\hs\epsilon_{(-1,1)} \Bigr) .
\end{equation}
Unlike the free theory, the interacting theory does not have a variation of the gauge parameters which keeps
$\delta_0\Phi$ strictly invariant. Eq.~\eqref{tQez} tells us, however, that the variation 
\begin{equation} \label{del1:Berkovits:full}
\delta_1
\bbm
\epsilon_{(-1,0)} \\[1ex]
\epsilon_{(-1,1)}
\ebm
= 
\bbm
Q&\e^{\adg}\ez &0 \\[.5ex]
0&\tQ&\ez
\ebm
\bbm
\epsilon_{(-2,0)} \\[.2ex]
\epsilon_{(-2,1)} \\[.2ex]
\epsilon_{(-2,2)}
\ebm
\end{equation}
does not change \eqref{delP} if we use the equation of motion
\begin{equation} \label{eom:B}
\frac{\delta\SB}{\delta G} \left(\propto \frac{\delta\SB}{\delta\Phi}\right)=0\,.
\end{equation}
Indeed, noting
\bs
\begin{align}
\delta_1\bigl( Q\epsilon_{(-1,0)}\bigr) &= Q\bigl(\delta_1\epsilon_{(-1,0)}\bigr) 
= Q \e^{\adg}\ez \epsilon_{(-2,1)}\,,
\label{dQe} \\[.5ex]
\delta_1\bigl( \ez\epsilon_{(-1,1)}\bigr) &= \ez\bigl(\delta_1\epsilon_{(-1,1)}\bigr)
= \ez\tQ\epsilon_{(-2,1)}\,,
\label{deze}
\end{align}
\es
we have
\begin{align} \label{Berkovits:d_1:d_0}
\delta_1\bigl( \delta_0 \Phi\bigr)
&= \frac{\adg}{1-\e^{-\adg}}\, \Bigl( \e^{-\adg} \delta_1\bigl( Q\epsilon_{(-1,0)}\bigr) + \delta_1\bigl(\ez\epsilon_{(-1,1)}\bigr) \Bigr) 
\nonumber \\[.5ex]
&= \frac{\adg}{1-\e^{-\adg}}\, \bigl\{ \tQ, \ez\bigr\} \epsilon_{(-2,1)}\,.
\end{align}
Furthermore, eq.~\eqref{del1:Berkovits:full} is invariant under the variation
\begin{equation}
\delta_2 
\bbm
\epsilon_{(-2,0)} \\[.2ex]
\epsilon_{(-2,1)} \\[.2ex]
\epsilon_{(-2,2)}
\ebm
=
\bbm
Q&\e^{\adg}\ez &0 &0\\[.5ex]
0&\tQ&\ez &0 \\[.5ex]
0&0&\tQ&\ez
\ebm
\bbm
\epsilon_{(-3,0)} \\[.2ex]
\epsilon_{(-3,1)} \\[.2ex]
\epsilon_{(-3,2)} \\[.2ex]
\epsilon_{(-3,3)}
\ebm
\end{equation}
if we use \eqref{eom:B}.
In fact, if the equation of motion is satisfied,
the $n$-th variation in the interacting theory
\begin{equation} \label{deln:Berkovits:full}
\delta_n
\begin{bmatrix}
\epsilon_{(-n, 0)} \\
\vdots \\
\epsilon_{(-n, n)}
\end{bmatrix}
= \ n+1 \underbrace{\left\{
\begin{bmatrix}
Q&\e^{\adg}\ez &&& \hsymbu{0}\!\!\!\!\!\\[.5ex]
&\tQ & \eta_0 & &  \\[.5ex]
&& \ddots & \,\ddots & \\[.5ex]
\hsymbl{0} && & \tQ & \quad\eta_0 \\
\end{bmatrix}
\right.}_{n+2}
\begin{bmatrix}
\epsilon_{(-(n+1), 0)} \\
\vdots \\
\epsilon_{(-(n+1), n+1)}
\end{bmatrix} 
\quad ( n\geq 1)
\end{equation}
is not affected by the $(n+1)$-st variation of the form
\begin{equation}
\delta_{n+1}
\begin{bmatrix}
\epsilon_{(-(n+1), 0)} \\
\vdots \\
\epsilon_{(-(n+1), n+1)}
\end{bmatrix}
= \ n+2 \underbrace{\left\{
\begin{bmatrix}
Q&\e^{\adg}\ez &&& \hsymbu{0}\!\!\!\!\!\\[.5ex]
&\tQ & \eta_0 & &  \\[.5ex]
&& \ddots & \,\ddots & \\[.5ex]
\hsymbl{0} && & \tQ & \quad\eta_0 \\
\end{bmatrix}
\right. }_{n+3}
\begin{bmatrix}
\epsilon_{(-(n+2), 0)} \\
\vdots \\
\epsilon_{(-(n+2), n+2)}
\end{bmatrix} 
.
\end{equation}
Note that in the first matrix on the right-hand side of \eqref{deln:Berkovits:full},
the first row contains $Q$ (not $\tQ$) and $\e^{\adg}\ez$, whereas the others contain $\tQ$ and $\ez$.

%--------------------------------------------------------------------------
\subsection{Relation between the reducibility structures of the two formulations}
\label{relation:reducibility}
In the preceding subsection, we have explained the reducibility structure of the Berkovits formulation before partial gauge fixing.
There, starting from \eqref{delP} with the two gauge parameters independent, 
we have obtained a series of the variations \eqref{deln:Berkovits:full}, which satisfy
\begin{equation}
\delta_{n+1}\left(\delta_n
\begin{bmatrix}
\epsilon_{(-n, 0)} \\
\vdots \\
\epsilon_{(-n, n)}
\end{bmatrix}
\right)
\simeq 0
\quad ( n\geq 0)\,.
\end{equation}
For example, the variations $\delta_1\epsilon_{(-1,0)}$ and $\delta_1\epsilon_{(-1,1)}$ are given by (see \eqref{del1:Berkovits:full})
\bs \label{del_1:ab}
\begin{align} 
\delta_1\epsilon_{(-1,0)} &= Q\epsilon_{(-2,0)} + \e^{\adg}\ez \epsilon_{(-2,1)}\,,
\label{del_1:a} \\[.5ex]
\delta_1\epsilon_{(-1,1)} &= \tQ\epsilon_{(-2,1)} + \ez \epsilon_{(-2,2)}\,.
\label{del_1:b}
\end{align}
\es
If once the partial gauge fixing is performed, however,
the parameters $\epsilon_{(-1,0)}$ and $\epsilon_{(-1,1)}$ are not independent any longer:
$\ez\hs\epsilon_{(-1,1)}$ can be expressed in terms of $Q\hs\epsilon_{(-1,0)}$ as in \eqref{recursion}.
Hence the reducibility structure is altered.
In the present subsection, we will investigate how it is altered by the partial gauge fixing,
and show that it includes, as a sub-structure, the regularized version of the reducibility structure of the Witten formulation.
Before beginning the analysis, let us summarize below the results to be obtained,
in order to clarify the direction in which we are going.
\begin{enumerate}
\item
The residual gauge transformation \eqref{explicit} is parameterized only by $\epsilon_{(-1,0)}$,
so that $\epsilon_{(-1,1)}$ disappears from the reducibility structure.
Then, this disappearance entails the disappearance of $\epsilon_{(-2,2)}$.
In fact, if a parameter $\epsilon_{(-n,n)}$ ($n\geq 1$) disappears from the reducibility structure,
so does $\epsilon_{(-(n+1),\,n+1)}$:
in \eqref{deln:Berkovits:full}, the parameter $\epsilon_{(-(n+1),\,n+1)}$ appears only in the variation
\begin{equation}
\delta_n\epsilon_{(-n,n)} = \tQ\epsilon_{(-(n+1),\,n)} + \ez \epsilon_{(-(n+1),\,n+1)}\,,
\end{equation}
and therefore the disappearance of $\epsilon_{(-n,n)}$ entails the disappearance of $\epsilon_{(-(n+1),\,n+1)}$.
Hence, under the partial gauge fixing \eqref{gfc}, 
all the parameters of the form $\epsilon_{(-n,n)}$ ($n\geq 1$) disappear from the reducibility structure.
Therefore, instead of \eqref{deln:Berkovits:full} we obtain
\begin{equation} 
\delta_n
\begin{bmatrix}
\epsilon_{(-n, 0)} \\
\vdots \\
\epsilon_{(-n, n-1)}
\end{bmatrix}
= \ n \underbrace{\left\{
\begin{bmatrix}
Q&\e^{\adg}\ez &&& \hsymbu{0}\!\!\!\!\!\\[.5ex]
&\tQ & \eta_0 & &  \\[.5ex]
&& \ddots & \ddots & \\[.5ex]
\hsymbl{0} && & \tQ & \quad\eta_0 \\
\end{bmatrix}
\right.}_{n+1}
\begin{bmatrix}
\epsilon_{(-(n+1), 0)} \\
\vdots \\
\epsilon_{(-(n+1), n)}
\end{bmatrix} 
\quad ( n\geq 1)\,.
\end{equation}
\item
Extending condition \eqref{gfc},
we consider the following restriction:
\begin{equation} \label{restriction:-n,0}
\Xil\hs\epsilon_{(-n,0)} = 0\qquad
(\forall n\geq 0\hs;\ \, \epsilon_{(0,0)} := \Phi)\,.
\end{equation}
Just as $\epsilon_{(-n,n)}$ ($n\geq 1$) disappear from the reducibility structure under condition \eqref{gfc},
many of the parameters $\epsilon_{(-n,m)}$ ($0\leq m\leq n$) disappear under the above restriction.
In fact, the reducibility sub-structure specified by \eqref{restriction:-n,0} is described only by $\epsilon_{(-n,0)}$
($n\geq 0$).
Expressing them as
\begin{equation} 
\epsilon_{(-n,0)}=(-1)^n\,\Xil\hs\Lambda_{(-(n-1),-1)}
\qquad \text{with}\qquad
\Lambda_{(-(n-1),-1)} \in\Hs\,,
\end{equation}
we find that $\Lambda_{(-(n-1),-1)}$ are counterparts of $\LW_{(-(n-1),-1)}$ in \eqref{deln:Witten:full}
and the sub-structure can be regarded as the regularized version of the reducibility structure of the Witten formulation.
\end{enumerate}

Now let us confirm the above.
We first investigate how the reducibility structure is altered by the partial gauge fixing \eqref{gfc}. 
For this purpose, it is convenient to start from the following form of the residual gauge transformation:
\begin{equation} \label{hd0}
\hd_0 \Phi
= \frac{\adg}{1-\e^{-\adg}}\, \Bigl( \e^{-\adg} Q\hs\epsilon_{(-1,0)} + \ez\hs\epsilon_{(-1,1)} \Bigr)\,,
\end{equation}
with $\ez\hs\epsilon_{(-1,1)}$ depending on $Q\hs\epsilon_{(-1,0)}$ as in \eqref{recursion}.
Here and in what follows, we append a hat to a variation symbol
when we consider a variation under the partial gauge fixing.
In order to find the variation $\hd_1\epsilon_{(-1,0)}$ which satisfies the reducibility relation
\begin{equation} \label{hd1:hd0}
\hd_1\bigl(\hd_0 \Phi\bigr) \simeq 0\,,
\end{equation}
let us try performing on $\epsilon_{(-1,0)}$ the transformation
\begin{equation} \label{hd1}
\hd_1 \epsilon_{(-1,0)} = Q\epsilon_{(-2,0)} + \e^{\adg}\ez\epsilon_{(-2,1)}
\end{equation}
as in \eqref{del_1:a}.
Then, through the relation \eqref{recursion}, $\ez\hs\epsilon_{(-1,1)}$ is transformed accordingly,
with its variation $\hd_1\bigl(\ez\hs\epsilon_{(-1,1)}\bigr)$ different from \eqref{deze}; 
hence $\hd_1\bigl(\hd_0 \Phi\bigr)$ does not coincide with \eqref{Berkovits:d_1:d_0}.
Nevertheless, eq.~\eqref{hd1} does lead to the desired relation \eqref{hd1:hd0}.
The point is that the variation $\hd_1\bigl(\ez\hs\epsilon_{(-1,1)}\bigr)$ induced by \eqref{hd1}
and the $\delta_1\bigl(\ez\hs\epsilon_{(-1,1)}\bigr)$ in \eqref{deze} is effectively the same:
\begin{equation} \label{hd1:Leta}
\hd_1\bigl(\ez\hs\epsilon_{(-1,1)}\bigr) \simeq \delta_1\bigl(\ez\hs\epsilon_{(-1,1)}\bigr)\,.
\end{equation}
In virtue of this relation, we obtain
\begin{equation}
\hd_1\bigl(\hd_0 \Phi\bigr)
= \frac{\adg}{1-\e^{-\adg}}\, \Bigl( \e^{-\adg} \hd_1\bigl( Q\epsilon_{(-1,0)}\bigr) + \hd_1\bigl(\ez\epsilon_{(-1,1)}\bigr) \Bigr) 
\simeq
\eqref{Berkovits:d_1:d_0} \simeq 0\,.
\end{equation}
We can indeed confirm \eqref{hd1:Leta} as follows, using eqs.~\eqref{recursion}, \eqref{hd1}, \eqref{deze}, and \eqref{tQez}:
\begin{align}
&\hd_1\bigl(\ez\hs\epsilon_{(-1,1)}\bigr) - \delta_1\bigl(\ez\hs\epsilon_{(-1,1)}\bigr) \nonumber \\
=& - \biggl[ 1 + \ez\Xil\biggl(\frac{-\adg}{\e^{-\adg}-1} -1\biggr)\biggr]^{-1}
\ez\Xil\,\frac{\adg}{\e^{\adg}-1}\,Q\e^{\adg}\ez\epsilon_{(-2,1)} -\ez\tQ\epsilon_{(-2,1)} \nonumber \\
=& - \biggl[ 1 + \ez\Xil\biggl(\frac{-\adg}{\e^{-\adg}-1} -1\biggr)\biggr]^{-1} \ez\Xil\,\frac{-\adg}{\e^{-\adg}-1}
\tQ\ez\epsilon_{(-2,1)} -\ez\tQ\epsilon_{(-2,1)} \nonumber \\
\simeq & - \biggl[ 1 + \ez\Xil\biggl(\frac{-\adg}{\e^{-\adg}-1} -1\biggr)\biggr]^{-1} \ez\Xil\,\frac{-\adg}{\e^{-\adg}-1}
\tQ\ez\epsilon_{(-2,1)} +\tQ\ez\epsilon_{(-2,1)} \nonumber \\
= &\ \biggl[ 1 + \ez\Xil\biggl(\frac{-\adg}{\e^{-\adg}-1} -1\biggr)\biggr]^{-1} \Xil\ez\tQ\ez\epsilon_{(-2,1)} \nonumber \\
\simeq &\ 0\qquad \Bigl(\because \ez\tQ\ez\simeq -\tQ\ez^2 = 0\Bigr) .
\end{align}
We thus find that the variation \eqref{hd1} is of an appropriate form.
Because the parameters $\epsilon_{(-2,0)}$ and $\epsilon_{(-2,1)}$ in \eqref{hd1} are independent,
we can continue the analysis on the reducibility structure in exactly the same manner as in subsection~\ref{red:Berkovits}.
The result is that the expression of the $n$-th variation is the same as \eqref{deln:Berkovits:full}
except that the parameters $\epsilon_{(-n,n)}$ and $\epsilon_{(-(n+1),n+1)}$ do not appear:  
\begin{equation} \label{hdn}
\hd_n
\begin{bmatrix}
\epsilon_{(-n, 0)} \\
\vdots \\
\epsilon_{(-n, n-1)}
\end{bmatrix}
= \ n \underbrace{\left\{
\begin{bmatrix}
Q&\e^{\adg}\ez &&& \hsymbu{0}\!\!\!\!\!\\[.5ex]
&\tQ & \eta_0 & &  \\[.5ex]
&& \ddots & \ddots & \\[.5ex]
\hsymbl{0} && & \tQ & \quad\eta_0 \\
\end{bmatrix}
\right.}_{n+1}
\begin{bmatrix}
\epsilon_{(-(n+1), 0)} \\
\vdots \\
\epsilon_{(-(n+1), n)}
\end{bmatrix} 
\quad ( n\geq 1)\,.
\end{equation}
Thus all the parameters of the form $\epsilon_{(-n,n)}$ ($n\geq 1$) has disappeared from the reducibility structure,
and we have
\begin{equation}
\hd_{n+1}\left(\hd_n
\begin{bmatrix}
\epsilon_{(-n, 0)} \\
\vdots \\
\epsilon_{(-n, n-1)}
\end{bmatrix}
\right)
\simeq 0
\quad ( n\geq 0)\,.
\end{equation}

Having examined the reducibility structure under the partial gauge fixing,
let us next show that it includes, as a sub-structure, the regularized version of the reducibility structure of the Witten formulation.
For this purpose, we restrict $\epsilon_{(-1,0)}$ within the subspace where \eqref{Xil:LQ} holds, as we did in subsection~\ref{residual symm}.
In fact, as will be explained, we can impose the same restriction on all the parameters of the form $\epsilon_{(-n,0)}$ ($n\geq 1$).
Just as the condition \eqref{gfc} for partial gauge fixing entails the elimination of $\epsilon_{(-n,n)}$ ($\forall n\geq 1$) 
from the set of independent parameters describing the reducibility structure,
the conditions $\Xil\epsilon_{(-1,0)}=0$, $\Xil\epsilon_{(-2,0)}=0$, $\Xil\epsilon_{(-3,0)}=0$, and so forth
eliminates respectively the parameters $\epsilon_{(-n,n-1)}$ ($\forall n\geq 2$), $\epsilon_{(-n,n-2)}$ ($\forall n\geq 3$),
$\epsilon_{(-n,n-3)}$ ($\forall n\geq 4$), and so forth.
Then what remains in the end is $\epsilon_{(-n,0)}$ ($n\geq 0$) with $\Xil\epsilon_{(-n,0)}=0$.
Expressing $\epsilon_{(-n,0)}$ as $\epsilon_{(-n,0)}=(-1)^n\Xil\Lambda_{(-(n-1),-1)}$, we find that 
$\Lambda_{(-(n-1),-1)}$ are the counterparts of $\LW_{(-(n-1),-1)}$ in the Witten formulation.
In the rest of the present section, we will confirm this argument.

First, let us investigate the consequence of the condition 
\begin{equation} \label{Xil:-1}
\Xil\hs\epsilon_{(-1,0)} = 0\,.
\end{equation}
Under this condition, the parameter $\epsilon_{(-2,1)}$ in \eqref{hd1} is not independent of $\epsilon_{(-2,0)}$:
in order for the transform of $\epsilon_{(-1,0)}$ to stay in the space in which \eqref{Xil:-1} holds,
the variation \eqref{hd1} has to satisfy
\begin{equation} \label{const:hd1}
\Xil\bigl(\hs\hd_1\epsilon_{(-1,0)}\bigr) \equiv \Xil\Bigl( Q\epsilon_{(-2,0)} + \e^{\adg}\ez\epsilon_{(-2,1)}\Bigr) = 0\,.
\end{equation}
This equation can be solved as follows, in the same manner that we used to obtain \eqref{recursion}.
From \eqref{const:hd1}, we have
\begin{equation}
\ez\epsilon_{(-2,1)} = \ez\Xil\Bigl(\ez\epsilon_{(-2,1)} - \hd_1\epsilon_{(-1,0)}\Bigr)
= -\ez\Xil\Bigl[Q\epsilon_{(-2,0)} + \bigl(\e^{\adg}-1\bigr)\ez\epsilon_{(-2,1)}\Bigr]\,.
\end{equation}
Solving this equation recursively, we obtain
\begin{align} \label{recursion:-2}
\ez\epsilon_{(-2,1)} 
&= -\sum^\infty_{n=0} \Bigl[-\ez\Xil\bigl(\e^{\adg}-1\bigr)\Bigr]^n\ez\Xil Q\epsilon_{(-2,0)} \nonumber \\
&= -\Bigl[1+ \ez\Xil\bigl(\e^{\adg}-1\bigr)\Bigr]^{-1} \ez\Xil Q\epsilon_{(-2,0)}\,.
\end{align}
In the last equality, we have used the fact that the sum in \eqref{recursion:-2} converges for small $g$ because the operator
\begin{equation}
-\ez\Xil\bigl(\e^{\adg}-1\bigr)
\end{equation}
is $O(g)$.
We thus find that $\ez\epsilon_{(-2,1)}$ can be expressed in terms of $Q\epsilon_{(-2,0)}$,
and that the explicit form of \eqref{hd1} under the constraint \eqref{const:hd1} is given by
\begin{equation} \label{explicit:-1}
\hdsub_1\epsilon_{(-1,0)} 
= Q\epsilon_{(-2,0)} + \e^{\adg}\ez\epsilon_{(-2,1)}
= \biggl( 1- \e^{\adg} \Bigl[1+ \ez\Xil\bigl(\e^{\adg}-1\bigr)\Bigr]^{-1} \ez\Xil\biggr) Q\epsilon_{(-2,0)}\,.
\end{equation}
Here and in what follows, the ``sub'' on the variation symbol indicates that we consider the reducibility sub-structure
specified by \eqref{restriction:-n,0}.
In order to find the variation $\hdsub_2\epsilon_{(-2,0)}$ which satisfies the reducibility relation
\begin{equation} \label{hdsub2:hdsub1}
\hdsub_2\bigl(\hs\hdsub_1\epsilon_{(-1,0)}\bigr) \simeq 0\,,
\end{equation}
let us try performing on $\epsilon_{(-2,0)}$ the transformation
\begin{equation} \label{hd2:-2}
\hdsub_2 \epsilon_{(-2,0)} = Q\epsilon_{(-3,0)} + \e^{\adg}\ez\epsilon_{(-3,1)}
\end{equation}
as in \eqref{hdn}.
The point is that just as \eqref{hd1:Leta} holds, the variation $\hdsub_2\bigl(\ez\epsilon_{(-2,1)}\bigr)$ 
induced by \eqref{hd2:-2} through the relation \eqref{recursion:-2} is effectively the same as 
$\hd_2\bigl(\ez\epsilon_{(-2,1)}\bigr)$, which is obtained directly from the equation
\begin{equation}
\hd_2 \epsilon_{(-2,1)} = \tQ \epsilon_{(-3,1)} + \ez \epsilon_{(-3,2)}
\end{equation}
in \eqref{hdn}:
\begin{equation} \label{hd2ind}
\hdsub_2 \bigl(\ez\epsilon_{(-2,1)}\bigr) \simeq \hd_2\bigl(\ez\epsilon_{(-2,1)}\bigr).
\end{equation}
The proof of \eqref{hd2ind} goes along the same lines as that of \eqref{hd1:Leta}.
It follows from \eqref{hd2ind} that we have
\begin{align}
\hdsub_2\bigl(\hs\hdsub_1\epsilon_{(-1,0)}\bigr)
&= \hdsub_2\bigl(Q\epsilon_{(-2,0)}\bigr) + \e^{\adg}\hdsub_2\bigl(\ez\epsilon_{(-2,1)}\bigr)
\nonumber \\[.5ex]
&\simeq \hdsub_2\bigl(Q\epsilon_{(-2,0)}\bigr) + \e^{\adg}\hd_2\bigl(\ez\epsilon_{(-2,1)}\bigr)
= \e^{\adg} \{ \tQ, \ez\} \hs\epsilon_{(-3,1)} \simeq 0\,.
\end{align}
Thus \eqref{hd2:-2} provides the desired relation \eqref{hdsub2:hdsub1}
even under the constraint \eqref{Xil:-1}.
In this manner, the parameters $\epsilon_{(-2,1)}$ and $\epsilon_{(-3,2)}$ have vanished from 
the set of independent parameters describing the reducibility structure up to this stage.
As can be seen from \eqref{hdn}, 
this vanishment entails the vanishment of $\epsilon_{(-4,3)}$, $\epsilon_{(-5,4)}$, $\epsilon_{(-6,5)}$, and so forth.
Indeed, in \eqref{hdn}, a parameter $\epsilon_{(-(n+1),\,n)}$ appears only in the variation
\begin{equation}
\hd_n\epsilon_{(-n,n-1)} = \tQ\epsilon_{(-(n+1),\,n-1)} + \ez \epsilon_{(-(n+1),\,n)}\,,
\end{equation}
and therefore if $\epsilon_{(-n,n-1)}$ ($n\geq 2$) vanishes from the reducibility structure, so does $\epsilon_{(-(n+1),\,n)}$.
Hence, as a consequence of condition \eqref{Xil:-1},
all the parameters of the form $\epsilon_{(-n,n-1)}$ ($\forall n\geq 2$) disappear from the reducibility structure.

Next let us consider the condition
\begin{equation} \label{cond:-2}
\Xil\hs\epsilon_{(-2,0)} = 0\,,
\end{equation}
which can be realized by the use of the degree of freedom of the transformation \eqref{hd2:-2}.
Under this condition, the parameters $\epsilon_{(-3,0)}$ and $\epsilon_{(-3,1)}$ in \eqref{hd2:-2} are not independent any longer.
Because the structure of \eqref{hd2:-2} is exactly the same as that of \eqref{hd1}, we can express $\ez\epsilon_{(-3,1)}$
in terms of $Q\epsilon_{(-3,0)}$, following the same argument as before.
This time, through the condition \eqref{cond:-2}, the parameters $\epsilon_{(-3,1)}$, $\epsilon_{(-4,2)}$, $\epsilon_{(-5,3)}$, and so forth are
eliminated from the set of independent parameters.
We can proceed our argument in this manner, and in the end we obtain the reducibility sub-structure described only by 
the parameters $\epsilon_{(-n,0)}$ ($n\geq 1$) with 
\begin{equation} \label{const:Xil:-n,0}
\Xil\epsilon_{(-n,0)} = 0\quad (\forall n\geq 1)\,.
\end{equation}
The variations of these parameters take the same form as \eqref{explicit:-1}:
\begin{align} \label{explicit:-n}
\hdsub_n\epsilon_{(-n,0)} &= \biggl( 1- \e^{\adg} \Bigl[1+ \ez\Xil\bigl(\e^{\adg}-1\bigr)\Bigr]^{-1} \ez\Xil\biggr) Q\epsilon_{(-(n+1),0)}
\nonumber \\[.5ex]
&= -\Xil Q\ez\epsilon_{(-(n+1),0)} -g\hs\Xil\Bigl[\ez\Phi\,,\, \ez\Xil Q\epsilon_{(-(n+1),0)}\Bigr\} + O(g^2)\,,
\end{align}
with
\begin{align}
\hdsub_{n+1} \bigl(\hs \hdsub_n\epsilon_{(-n,0)} \bigr) \simeq 0 \qquad (n\geq 0)\,.
\end{align}
In virtue of the relation \eqref{const:Xil:-n,0},
the parameters $\epsilon_{(-n,0)}$ can be expressed as 
\begin{equation} \label{en}
\epsilon_{(-n,0)}=(-1)^n\,\Xil\hs\Lambda_{(-(n-1),-1)}
\end{equation}
for some $\Lambda_{(-(n-1),-1)}\in\Hs$.
Substituting \eqref{Phi:Xil:Psi} and \eqref{en} into \eqref{explicit:-n}, we obtain
\begin{equation} \label{n-th}
\hdsub_n\Lambda_{(-(n-1),-1)}
= Q\Lambda_{(-n,-1)} + g\bigl[ \Psi, \Xl \Lambda_{(-n,-1)}\bigr\} + O(g^2)\,.
\end{equation}
In the singular limit $\lambda\to 0$, eq.~\eqref{n-th} coincides with \eqref{deln:Witten:full} except for the $O(g^2)$ term,
under the identification
\begin{equation}
\Psi\cong\PsiW_{(1,-1)}\,,\quad 
\Lambda_{(-n,-1)} \cong \LW_{(-n,-1)}\,.
\end{equation}
Because the reducibility sub-structure \eqref{explicit:-n} of the Berkovits formulation
has no singularity for $\lambda\neq 0$,
we conclude that it can be regarded as the regularized version of the reducibility structure of the Witten formulation.

%%%%%%%%%%%%%%%%%%%%%%%%%%%%%%%%%%%%%%%%%%%%%%%%%%%%%%%
\section{Master action in the Batalin-Vilkovisky formalism}
\label{sec:master action}
\setcounter{equation}{0}
Quantization of complicated gauge systems such as string field theory is often performed 
with the Batalin-Vilkovisky (BV) formalism~\cite{BV1, BV2, Henneaux, GPS},
which is an extension of the BRST formalism.
In this formalism, the most important process for gauge fixing is construction of the master action, 
or the solution to the classical master equation.
The equation is an extension of the Ward-Takahashi identity,
and the point is that given a reducibility structure, we can in principle construct its solution.
Taking this into account, we expect from the result in subsection~\ref{relation:reducibility}  
that the master action in the Berkovits formulation will be related to that in the Witten formulation.
In the present section, we will show that it is indeed the case: 
the former reduces to the regularized version of the latter after partial gauge fixing.

Because superstring field theory in the Witten formulation has the same structure as bosonic string field theory~\cite{Witten:bosonic},
apart from the problem of the picture-changing operator, we can easily obtain its master action $\calSW$ formally
as in the bosonic case~\cite{Thorn}.
It is given by
\begin{equation} \label{Witten:master action}
\calSW = -\frac{1}{2} \bllangle\bPsiW , Q\bPsiW \brrangle 
-\frac{g}{3} \bllangle \bPsiW, \Xmid\bigl(\bPsiW\ast\bPsiW\bigr) \brrangle \,,
\end{equation}
where
\begin{equation}
\bPsiW = \sum^\infty_{\boldsymbol{g} =-\infty} \PsiW_{(\boldsymbol{g},-1 )}\,.
\end{equation}
Here the $\PsiW_{(\boldsymbol{g}, -1)}$ with $\boldsymbol{g}\leq 0$ are ghost fields,
and those with $\boldsymbol{g}\geq 2$  are antighost fields.\footnote{
In the present paper, we will not distinguish antighosts and antifields in the BV formalism for simplicity.
In the language of the BV formalism, we will consider only a gauge-fixing fermion such that 
antifields of minimal-sector fields are identified with antighosts.
}
All of these $\Psi$'s are Grassmann odd.
Note that if we neglect the divergences caused by the picture-changing operator, the above action \eqref{Witten:master action} is indeed
a solution to the classical master equation of the form
\begin{equation}
\sum_{n\leq 1} \Bgllangle \frac{\delta_R \calSW}{\delta \PsiW_{(n, -1)}}\,,\, \frac{\delta_L \calSW}{\delta \PsiW_{(3-n ,-1)}} \Bgrrangle
= 0\,,
\end{equation}
where $\delta_R$ and $\delta_L$ denote the right and the left variation, respectively.
We will demonstrate that the master action in the Berkovits formulation 
reduces to the regularized version of \eqref{Witten:master action}
after we perform partial gauge fixing and integrate out auxiliary components.

%-------------------------------------------------------------
\subsection{Relation between the master actions in the free theories}
Let us begin by considering the free theories, in which the coupling $g$ is equal to zero.
In this case, the master action \eqref{Witten:master action} in the Witten formulation becomes
\begin{equation} \label{Witten:ma:free}
\calSW_\mrm{quad} := \calSW \bigl|_{g=0}\; =-\frac{1}{2} \bllangle\bPsiW , Q\bPsiW \brrangle 
\,,
\qquad
\bPsiW = \sum^\infty_{\boldsymbol{g} =-\infty} \PsiW_{(\boldsymbol{g},-1 )}\,.
\end{equation}
Noting eq.~\eqref{relation:small}, we can rewrite this as
\begin{equation}
\calSW_\mrm{quad} = \sum^\infty_{n=0} \calSW_n\,,
\end{equation}
with
\bs
\begin{align}
\calSW_0 &= \SW \bigl|_{g=0}\; = -\frac{1}{2} \bllangle\PsiW_{(1,-1)} , Q\PsiW_{(1,-1)} \brrangle \,,
\\[2.5ex]
\calSW_n &= - \bllangle\PsiW_{(n+1,-1)} , Q\PsiW_{(-n+1,-1)} \brrangle
\qquad
(n \geq 1)\,. \label{SW_n}
\end{align}
\es
The free master action $\calSB_\mrm{quad}$ in the Berkovits formulation is given in ref.~\cite{paperI}:
\begin{equation} \label{Berkovits:ma:free}
\calSB_\mrm{quad} = \sum^\infty_{n=0} \calSB_n\,,
\end{equation}
with
\bs
\begin{align}
\calSB_0 &= \SB\bigl|_{g=0}\; = -\frac{\iu}{2}\, \bigl\langle \Phi_{(0,0)}, Q\ez\Phi_{(0,0)}\bigr\rangle\,,
\\[2.5ex]
\calSB_n &= \iu\sum^{n-1}_{m=0} \bigl\langle \Phi_{(n+1, -m-1)},\, Q\Phi_{(-n, m)} + \ez\Phi_{(-n, m+1)}\bigr\rangle
\qquad
(n \geq 1)\,. \label{SB_n}
\end{align}
\es
Here $\Phi_{(-n,m)}$ ($1\leq n$, $0\leq m \leq n$) are ghosts, and $\Phi_{(n+1,-m)}$ ($1\leq m \leq n$) are antighosts.
All the ghosts are Grassmann even, whereas all the antighosts are Grassmann odd.
The master action \eqref{Berkovits:ma:free} is a solution to the classical master equation in the Berkovits formulation 
of the form~\cite{Torii:proceedings}
\begin{equation} \label{Berkovits:me}
\sum^\infty_{n=0}\sum^n_{m=0} \Biggl\langle \frac{\delta_R \calSB}{\delta\Phi_{(-n,m)}}\,,\,
\frac{\delta_L \calSB}{\delta\Phi_{(n+2, -m-1)}}\Biggr\rangle = 0\,.
\end{equation}
Furthermore, as explained in ref.~\cite{Torii:validity}, it is invariant under the transformations bellow:
\bs
\begin{align}
\delta\Phi_{(-n,m)} &= Q\Lambda_{(-(n+1), m)} + \ez\Lambda_{(-(n+1), m+1)}\qquad (0\leq m\leq n)\,,\\[1ex]
\delta\Phi_{(2,-1)} &= Q\ez \Lambda_{(0,0)}\,,\\[1ex]
\delta\Phi_{(n+1, -1)} &= Q\Lambda_{(n, -1)} \qquad (2\leq n)\,,\\[1ex]
\delta\Phi_{(n+1, -m)} &= \ez\Lambda_{(n,-(m-1))} + Q\Lambda_{(n, -m)}\qquad (2\leq m\leq n-1)\,,\\[1ex]
\delta\Phi_{(n+1, -n)} &= \ez\Lambda_{(n, -(n-1))} \qquad (2\leq n)\,,
\end{align}
\es
where $\Lambda$'s are gauge parameters.
In the rest of the present subsection, we are going to show that $\calSB_\mrm{quad}$ reduces to $\calSW_\mrm{quad}$ under partial gauge fixing
for the above symmetry, which is an extension of the original gauge symmetry \eqref{del0:Phi}.\footnote{
Note that the relation between the completely gauge-fixed free actions have already been manifested in ref.~\cite{paperI}.
}
In fact, each $\calSB_n$ $(n\geq 0)$ reduces to $\calSW_n$\,.
For showing this, it is convenient to decompose $\Phi_{(\boldsymbol{g}, \boldsymbol{p})}$ as follows:
\begin{equation} \label{decomposition}
\Phi_{(\boldsymbol{g}, \boldsymbol{p})} = \{\ez, \Xil\}\hs \Phi_{(\boldsymbol{g}, \boldsymbol{p})} 
= \PhiN_{(\boldsymbol{g}, \boldsymbol{p})} + \Xil \PhiXi_{(\boldsymbol{g}, \boldsymbol{p})}\,,
\end{equation}
with
\begin{equation}
\PhiN_{(\boldsymbol{g}, \boldsymbol{p})} = \ez\Xil \Phi_{(\boldsymbol{g}, \boldsymbol{p})}\,,
\qquad
\PhiXi_{(\boldsymbol{g}, \boldsymbol{p})} = \ez \Phi_{(\boldsymbol{g}, \boldsymbol{p})}\,.
\end{equation}
The string fields $\PhiN_{(\boldsymbol{g}, \boldsymbol{p})}$ and $\PhiXi_{(\boldsymbol{g}, \boldsymbol{p})}$ are 
in the small Hilbert space.
Note that the subscript $(\boldsymbol{g}, \boldsymbol{p})$ on them is simply carried over from $\Phi_{(\boldsymbol{g}, \boldsymbol{p})}$
and does not indicate the ghost numbers and the picture numbers of 
$\PhiN_{(\boldsymbol{g}, \boldsymbol{p})}$ and $\PhiXi_{(\boldsymbol{g}, \boldsymbol{p})}$.
We impose the following conditions for partial gauge fixing:
\bs \label{pgf}
\begin{align}
\Xil \Phi_{(-n,m)} &= 0\qquad (0\leq m\leq n)\,,\\[1ex]
\Xil \Phi_{(n+1, -m)} &=0\qquad (2\leq m\leq n)\,,
\end{align}
\es
or equivalently
\bs \label{pgf:comp}
\begin{align}
\Phi_{(-n,m)} &= \Xil\PhiXi_{(-n,m)}\qquad (0\leq m\leq n)\,,
\label{pgf:comp:ghosts}\\[1ex]
\Phi_{(n+1, -m)} &= \Xil \PhiXi_{(n+1, -m)} \qquad (2\leq m\leq n)\,.
\label{pgf:comp:antighost}
\end{align}
\es
It should be noted that this set of conditions coincides with 
a subset of the conditions for complete gauge fixing of the free master action considered in subsection 2.2 of ref.~\cite{paperI},
merely by replacing $\Xil$ in \eqref{pgf} with $\xz$.

We have already learned in subsection~\ref{subsec:idea} that $\calSB_0$ reduces to $\calSW_0$ under the condition
\begin{equation} \label{cond:0,0}
\Xil\Phi_{(0,0)} = 0\,.
\end{equation}
Therefore, what we have to show in the present subsection is the reduction of $\calSB_n$ to $\calSW_n$ for $n\geq 1$.
Let us first consider the action
\begin{equation} \label{SB_1}
\calSB_1 = \iu\,\bigl\langle \Phi_{(2, -1)},\, Q\Phi_{(-1, 0)} + \ez\Phi_{(-1, 1)}\bigr\rangle
= \iu\,\bigl\langle \Phi_{(2, -1)},\, Q\Phi_{(-1, 0)} \bigr\rangle
+ \iu\,\bigl\langle \Phi_{(2, -1)},\, \ez\Phi_{(-1, 1)}\bigr\rangle\,.
\end{equation}
As can be seen from \eqref{pgf}, 
the field $\Phi_{(2,-1)}$ does not submit to any constraints, whereas $\Phi_{(-1,0)}$ and $\Phi_{(-1,1)}$ are subject to the conditions
\begin{equation} \label{pgf:n=1}
\Xil\Phi_{(-1,0)} = 0\,,\qquad
\Xil\Phi_{(-1,1)} = 0\,.
\end{equation}
Noting \eqref{prod=0}, we find that the second term on the rightmost side of \eqref{SB_1} becomes
\begin{equation}
\iu\,\bigl\langle \Phi_{(2, -1)},\, \ez\Phi_{(-1, 1)}\bigr\rangle
= \iu\,\bigl\langle \Xil\PhiXi_{(2, -1)},\, \PhiXi_{(-1, 1)}\bigr\rangle\,.
\end{equation}
Because $\PhiXi_{(-1, 1)}$ appears only in this term, it acts as a Lagrange multiplier field which imposes
\begin{equation}
\PhiXi_{(2, -1)} = 0\,.
\end{equation}
After integrating out $\PhiXi_{(-1, 1)}$, the action $\calSB_1$ therefore reduces to
\begin{align} \label{reduced:SB_1}
\calSB_1 &= \iu\,\bigl\langle \PhiN_{(2, -1)},\, Q\Xil\PhiXi_{(-1, 0)} \bigr\rangle
= \iu\,\bigl\langle \PhiN_{(2, -1)},\, \Xl\PhiXi_{(-1, 0)} \bigr\rangle
- \iu\,\bigl\langle \PhiN_{(2, -1)},\, \Xil Q\PhiXi_{(-1, 0)} \bigr\rangle\,.
\nonumber \\[1ex]
&= - \iu\,\bigl\langle \PhiN_{(2, -1)},\, \Xil Q\PhiXi_{(-1, 0)} \bigr\rangle
= \bllangle \PhiN_{(2, -1)},\, Q\PhiXi_{(-1, 0)} \brrangle\,.
\end{align}
In the third equality, we have used \eqref{prod=0}: both $\PhiN_{(2, -1)}$ and $\Xl\PhiXi_{(-1, 0)}$
are in the small Hilbert space (note \eqref{ezXl=0}), and therefore 
$\bigl\langle \PhiN_{(2, -1)},\, \Xl\PhiXi_{(-1, 0)} \bigr\rangle$ is zero.
The action \eqref{reduced:SB_1} thus coincides with $\calSW_1$ under the identification
\begin{equation}
\PhiN_{(2, -1)} \cong -\PsiW_{(2,-1)}\,,\qquad
\PhiXi_{(-1, 0)} \cong \PsiW_{(0,-1)}\,.
\end{equation}

Next we consider the action $\calSB_2$.
It takes the form
\begin{align} \label{SB_2}
\calSB_2 &= \iu\,\bigl\langle \Phi_{(3, -1)},\, Q\Phi_{(-2, 0)} + \ez\Phi_{(-2, 1)}\bigr\rangle
+ \iu\,\bigl\langle \Phi_{(3, -2)},\, Q\Phi_{(-2, 1)} + \ez\Phi_{(-2, 2)}\bigr\rangle
\nonumber \\[1ex]
&= \iu\,\bigl\langle \Phi_{(3, -1)},\, Q\Phi_{(-2, 0)} + \ez\Phi_{(-2, 1)}\bigr\rangle
+ \iu\,\bigl\langle \Phi_{(3, -2)},\, Q\Phi_{(-2, 1)} \bigr\rangle
+ \iu\,\bigl\langle \Phi_{(3, -2)},\, \ez\Phi_{(-2, 2)}\bigr\rangle\,.
\end{align}
The field $\Phi_{(3,-1)}$ does not obey any conditions, but the others do:
\bs \label{pgf:n=2}
\begin{align}
\Xil\Phi_{(-2,0)} &= 0\,,\quad
\Xil\Phi_{(-2,1)} = 0\,\,,\quad
\Xil\Phi_{(-2,2)} = 0\,\,, \label{pgf:n=2:ghost} \\[1ex]
\Xil\Phi_{(3,-2)} &= 0\,. \label{pgf:n=2:antighost}
\end{align}
\es
Noting \eqref{prod=0}, we realize that the last term on the rightmost side of \eqref{SB_2} becomes
\begin{equation}
\iu\,\bigl\langle \Phi_{(3, -2)},\, \ez\Phi_{(-2, 2)}\bigr\rangle
= \iu\,\bigl\langle \Xil\PhiXi_{(3, -2)},\, \PhiXi_{(-2, 2)}\bigr\rangle\,.
\end{equation}
Because $\PhiXi_{(-2, 2)}$ appears only in this term, it acts as a Lagrange multiplier field imposing
\begin{equation}
\PhiXi_{(3, -2)} =0\,.
\end{equation}
This constraint, together with \eqref{pgf:n=2:antighost}, means that $\Phi_{(3, -2)}$ should vanish:
\begin{equation}
\Phi_{(3, -2)} =0\,.
\end{equation}
Therefore, after we integrate out $\PhiXi_{(-2, 2)}$, the action $\calSB_2$ reduces to
\begin{align} \label{SB_2:reduced}
\calSB_2 = \iu\,\bigl\langle \Phi_{(3, -1)},\, Q\Phi_{(-2, 0)} + \ez\Phi_{(-2, 1)}\bigr\rangle
= \iu\,\bigl\langle \Phi_{(3, -1)},\, Q\Phi_{(-2, 0)} \bigr\rangle
+ \iu\,\bigl\langle \Phi_{(3, -1)},\, \ez\Phi_{(-2, 1)}\bigr\rangle\,.
\end{align}
This is similar in form to \eqref{SB_1}:
the fields $\Phi_{(3, -1)}$, $\Phi_{(-2, 0)}$, and $\Phi_{(-2, 1)}$ in \eqref{SB_2:reduced} correspond to
$\Phi_{(2, -1)}$, $\Phi_{(-1, 0)}$, and $\Phi_{(-1, 1)}$ in \eqref{SB_1}, respectively.
Consequently, the argument goes along the same lines as before:
$\PhiXi_{(-2,1)}$ acts as a Lagrange multiplier which imposes
\begin{equation}
\PhiXi_{(3,-1)} =0\,,
\end{equation}
and after integrating out $\PhiXi_{(-2,1)}$ we obtain
\begin{equation} \label{SB_2:reduced2}
\calSB_2 = \bllangle \PhiN_{(3, -1)},\, Q\PhiXi_{(-2, 0)} \brrangle\,.
\end{equation}
This action coincides with $\calSW_2$ under the identification
\begin{equation}
\PhiN_{(3, -1)} \cong -\PsiW_{(3,-1)}\,,\qquad
\PhiXi_{(-2, 0)} \cong \PsiW_{(-1,-1)}\,.
\end{equation}
The process of the reduction of $\calSB_2$ to $\calSW_2$ can be summarized as follows.
First, $\PhiXi_{(-2,2)}$ acts as the Lagrange multiplier which imposes $\PhiXi_{(3,-2)} =0$.
Integrating out this field, we obtain the reduced action which has the same structure as $\calSB_1$.
Then, $\PhiXi_{(-2,1)}$ acts as the second Lagrange multiplier which imposes $\PhiXi_{(3,-1)} =0$,
and finally we obtain the completely reduced action \eqref{SB_2:reduced2}.
In fact, for the case of $\calSB_n$ ($n\geq 1$), the situation is the same:
The field $\PhiXi_{(-n,n)}$ in \eqref{SB_n} acts as the Lagrange multiplier imposing $\PhiXi_{(n+1,-n)} =0$.
We first integrate out this field, and obtain the reduced action of the same structure as $\calSB_{n-1}$.
Then in this reduced action, $\PhiXi_{(-n,n-1)}$ acts as the Lagrange multiplier imposing $\PhiXi_{(n+1,-(n-1))} =0$.
Integrating out that, we obtain the action of the same structure as $\calSB_{n-2}$.
Continuing this process, 
we find that the fields $\PhiXi_{(-n,n)}$, $\PhiXi_{(-n,n-1)}$, ..., $\PhiXi_{(-n,1)}$ are integrated out as Lagrange multipliers
eliminating $\PhiXi_{(n+1,-n)}$, $\PhiXi_{(n+1, -(n-1))}$, ..., $\PhiXi_{(n+1,-1)}$, respectively.
In the end, only the fields $\PhiXi_{(-n,0)}$ and $\PhiN_{(n+1,-1)}$ survive,
and we achieve the completely reduced action
\begin{equation} \label{SB_n:reduced}
\calSB_n = \bllangle \PhiN_{(n+1, -1)},\, Q\PhiXi_{(-n, 0)} \brrangle\,,
\end{equation}
which coincides with \eqref{SW_n} under the identification
\begin{equation} \label{identification}
\PhiN_{(n+1, -1)} \cong -\PsiW_{(n+1,-1)}\,,\qquad
\PhiXi_{(-n, 0)} \cong \PsiW_{(-n+1,-1)}\,.
\end{equation}

%-------------------------------------------------------------
\subsection{Relation between the master actions in the interacting theories}
Now that we have confirmed the correspondence of the master actions in the free theories,
let us next manifest the relation between the interacting theories.
Unlike the Witten theory, the Berkovits theory remains regular even if the interaction is present,
being free from the midpoint insertion of the picture-changing operator.
In fact, the master action in the Berkovits formulation, $\calSB$,
can be interpreted as the regularized version of that in the Witten formulation.
In particular, in the singular limit $\lambda \to 0$ of the partial gauge fixing \eqref{pgf},
the action $\calSB$, which is non-polynomial, 
reproduces the formal cubic master action \eqref{Witten:master action} up to $O(g^2)$ terms.
(The deviation will play the role of the counterterms for canceling the divergences in the Witten formulation.)
For showing this, it is sufficient to examine the cubic term $\calSB_\mrm{cubic}$ of the master action, 
which are of order $g$, as well as the quadratic term $\calSB_\mrm{quad}$.\footnote{
The complete form of the master action in the Berkovits formulation has not yet been obtained,
although several attempts have been made~\cite{paperII, constrainedBV}.
}

Before moving into the investigation of $\calSB_\mrm{cubic}$,
it is helpful to review what we have learned from the analysis of $\calSB_\mrm{quad}$ in the free theory.
In the preceding subsection, 
the field $\PhiXi_{(-n,m)}$ ($1\leq m\leq n$) in $\calSB_n$ acts as the $(n+1-m)$-th Lagrange multiplier imposing
\begin{equation} \label{constraint:PhiXi:n+1,-m}
\PhiXi_{(n+1,-m)} =0
\qquad (1\leq m\leq n)\,.
\end{equation}
After integrating out all of these Lagrange multipliers one by one, 
only the fields $\PhiXi_{(-n,0)}$ and $\PhiN_{(n+1,-1)}$ survive out of 
the ghosts $\Phi_{(-n,m)}$ ($1\leq n$, $0\leq m \leq n$) and the antighosts $\Phi_{(n+1,-m)}$ ($1\leq m \leq n$),
with the survivors corresponding to $\PsiW_{(-n+1,-1)}$ and $-\PsiW_{(n+1,-1)}$ in the Witten formulation.
As we will see, the interacting theory succeeds to this structure,
with the right-hand side of \eqref{constraint:PhiXi:n+1,-m} receiving $O(g)$ correction:
\begin{equation} \label{constraint:O(g)}
\PhiXi_{(n+1,-m)} =O(g)
\qquad (1\leq m\leq n)\,.
\end{equation}
In the process of the reduction of the master action $\calSB_\mrm{quad} + \calSB_\mrm{cubic}$, 
these $O(g)$ corrections include terms involving fields 
which do not appear in the completely reduced action in the free theory.
However, substituting the relations \eqref{constraint:O(g)} themselves back for these fields,
we will find that such extra terms are of order $g^2$,
and therefore they can be neglected in our analysis.
We will obtain in the end the completely reduced action described only by the fields 
$\PhiXi_{(0,0)}$, $\PhiXi_{(-n,0)}$ and $\PhiN_{(n+1,-1)}$ ($n\geq 1$) as in the free theory.
In the rest of the present subsection, we confirm what we have summarized above.

The complete form of the cubic terms of the master action in the Berkovits formulation, $\calSB_\mrm{cubic}$, 
was first shown in ref.~\cite{Torii:proceedings},
but there are many other expressions which are related to one another through canonical transformations in the BV formalism.\footnote{
The sum of $\calSB_\mrm{quad}$ and $\calSB_\mrm{cubic}$ satisfies the master equation \eqref{Berkovits:me} in the following sense:
\begin{equation*} 
\sum^\infty_{n=0}\sum^n_{m=0} \Biggl\langle \frac{\delta_R \bigl(\calSB_\mrm{quad}+\calSB_\mrm{cubic}\bigr)}{\delta\Phi_{(-n,m)}}\,,\,
\frac{\delta_L \bigl( \calSB_\mrm{quad}+\calSB_\mrm{cubic}\bigr)}{\delta\Phi_{(n+2, -m-1)}}\Biggr\rangle = O(g^2)\,.
\end{equation*}
}
They can in general be divided into four types of term as follows according to the numbers of $Q$ and $\ez$~\cite{paperII}:
\begin{equation}
\calSB_\mrm{cubic} = \calSB_{Q\eta} + \calSB_Q + \calSB_\eta\ + \calSB_N\,,
\end{equation}
where $\calSB_{Q\eta}$, $\calSB_Q$, $\calSB_\eta$, and $\calSB_N$ are the terms with one $Q$ and one $\ez$,
with one $Q$ and no $\ez$, with no $Q$ and one $\ez$, and with no $Q$ and no $\ez$, respectively.
In fact, $\calSB_{Q\eta}$ is exactly the cubic term in the original action $\SB$:
\begin{equation}
\calSB_{Q\eta} = \frac{\iu}{6}\,g\,\Bigl\langle \ez\Phi_{(0,0)}\,,\, \bigl[ \Phi_{(0,0)}, Q\Phi_{(0,0)}\bigr]\Bigr\rangle\,.
\end{equation}
Furthermore, the form of $\calSB_N$ is uniquely determined as
\begin{equation}
\calSB_N = \iu\, g\sum^\infty_{n_1, n_2 = 0}\ 
\sum_{\substack{0\leq m_1 \leq n_1 \\[.5ex] 0\leq m_2 \leq n_2}}
\Bigl\langle \Phi^\ast_{(n_1+2,\, -m_1 -1)} \Phi^\ast_{(n_2 +2,\, -m_2 -1)}\,,\, \Phi_{(-n_1 -n_2 -2,\, m_1 + m_2 +1)} \Bigr\rangle\,.
\end{equation}
(For the proof of the uniqueness, see ref.~\cite{paperII}.)
Here and in what follows, we append the superscript ``$\ast$'' to antighosts for convenience,
and consequently $\PhiXi_{(n+1,-m)}$ and $\PhiN_{(n+1,-m)}$ ($1\leq m\leq n$) will be written as $\Phi^{\ast\Xi}_{(n+1,-m)}$
and $\Phi^{\ast -}_{(n+1,-m)}$, respectively.
The degrees of freedom of canonical transformations in the BV formalism
are reflected in $\calSB_Q$ and $\calSB_\eta$.
Among many different expressions of $\calSB_Q$ and $\calSB_\eta$, 
in the present paper we will use the following one~\cite{paperII}:
\begin{align}
\calSB_Q 
=& -\iu\,g\sum^\infty_{n=0}\, \sum^n_{m=0} \Bigl\langle \Phi^\ast_{(n+2,\,-m-1)}\,,\, \Phi_{(-n-1,\,m)}\bigl(Q\Phi_{(0,0)}\bigr)\Bigr\rangle
\nonumber \\[.5ex]
&- \iu\,g\sum^\infty_{k=1}\, \sum^\infty_{a,b=0} \Bigl\langle \Phi^\ast_{(2+k+a+b,\,-1-k-b)}\,,\, \Phi_{(-1-a-b,\,b)}\bigl(Q\Phi_{(-k,\,k)}\bigr)\Bigr\rangle
\,,\\[1ex]
\calSB_\eta 
=& -\iu\,g\sum^\infty_{n=0}\, \sum^n_{m=0} \Bigl\langle \Phi^\ast_{(n+2,\,-m-1)}\,,\, \bigl(\ez\Phi_{(0,0)}\bigr)\Phi_{(-n-1,\,m+1)}\Bigr\rangle
\nonumber \\[.5ex]
&- \iu\,g\sum^\infty_{k=1}\, \sum^\infty_{a,b=0} \Bigl\langle \Phi^\ast_{(2+k+a+b,\,-1-b)}\,,\, \bigl(\ez\Phi_{(-k,0)}\bigr)\Phi_{(-1-a-b,\,1+b)}\Bigr\rangle
\,.\label{SB_eta}
\end{align}
The advantage of this choice is that $\calSB_\mrm{cubic}$ does not include
terms quadratic in an auxiliary field $\PhiXi_{(-n,m)}$ ($1\leq m\leq n$).
In fact, at any step of the reduction process, a descendant of $\calSB_\mrm{quad} + \calSB_\mrm{cubic}$
does not include such terms as long as we neglect $O(g^2)$ terms.
Therefore we can treat the auxiliary fields simply as Lagrange multipliers as in the free theory analysis.

Our strategy for investigating the relation between $\calSB$ and $\calSW$ 
under the conditions \eqref{pgf} (or equivalently \eqref{pgf:comp}) is as follows.
First, we pick up from $\calSB_\mrm{quad} + \calSB_\mrm{cubic}$ 
the terms including a first Lagrange multiplier field $\PhiXi_{(-n,n)}$ ($n\geq 1$),
which imposes a constraint on $\Phi^{\ast\Xi}_{(n+1,-n)}$,
and then integrate out all of these multipliers.
Next, from the resultant action,
we pick up the terms including a second Lagrange multiplier field $\PhiXi_{(-n,n-1)}$ ($n\geq 2$),
which imposes a constraint on $\Phi^{\ast\Xi}_{(n+1,-(n-1))}$,
and then integrate out these multipliers.
Continuing this process, we lastly obtain the completely reduced action of $\calSB$,
and compare it with $\calSW$.

Following the above strategy, 
let us begin by examining the terms including a first Lagrange multiplier field $\PhiXi_{(-n,n)}$ ($n\geq 1$):
\bs
\begin{align}
& \iu\,\Bigl\langle \Phi^\ast_{(n+1,-n)}, \ez\Phi_{(-n,n)}\Bigr\rangle\,,
\label{-n,n:quad}
\\[1ex]
& - \iu\,g \sum^\infty_{a,b=0} \Bigl\langle \Phi^\ast_{(2+n+a+b,\,-1-n-b)} \Phi_{(-1-a-b,\,b)}\,,\, Q\Phi_{(-n,n)} \Bigr\rangle\,,
\label{-n,n:Q}
\\[1ex]
& -\iu\,g \Bigl\langle \Phi^\ast_{(n+1,\,-n)} \bigl(\ez\Phi_{(0,0)}\bigr) \,,\, \Phi_{(-n,n)}\Bigr\rangle
- \iu\,g\sum^\infty_{k=1} \Bigl\langle \Phi^\ast_{(n+k+1,\,-n)} \bigl(\ez\Phi_{(-k,0)}\bigr) \,,\, \Phi_{(-n,n)}\Bigr\rangle\,,
\label{-n,n:eta}
\end{align}
\es
where \eqref{-n,n:quad}, \eqref{-n,n:Q}, and \eqref{-n,n:eta} 
are the contributions from $\calSB_\mrm{quad}$, $\calSB_Q$, and $\calSB_\eta$, respectively.
There are no contributions from $\calSB_{Q\eta}$ and $\calSB_N$.
Using the decomposition \eqref{decomposition}, the conditions \eqref{pgf}, and the BPZ evenness of $\Xil$,
we can express the sum of the above terms as
\begin{align}
& \iu\,\Bigl\langle \Xil\Phi^{\ast\Xi}_{(n+1,-n)}\,,\, \PhiXi_{(-n,n)}\Bigr\rangle
-\iu\,g\biggl\langle \Xil \!\!
\sum^\infty_{a,b=0} Q\Bigl( \Phi^\ast_{(2+n+a+b,\,-1-n-b)} \Phi_{(-1-a-b,\,b)}\Bigr) 
\,,\, \PhiXi_{(-n,n)} \biggr\rangle
\nonumber \\[.5ex]
& -\iu\,g\biggl\langle \Xil\biggl(
\Phi^\ast_{(n+1,\,-n)} \bigl(\ez\Phi_{(0,0)}\bigr)
+ \sum^\infty_{k=1} \Phi^\ast_{(n+k+1,\,-n)} \bigl(\ez\Phi_{(-k,0)}\bigr)
\biggr)\,,\, \PhiXi_{(-n,n)} \biggr\rangle\,.
\end{align}
Thus we find that $\PhiXi_{(-n,n)}$ imposes the constraint
\begin{align} \label{constraint:n+1,-n}
\Phi^{\ast\Xi}_{(n+1,-n)}
&= g\,\ez\Xil\Biggl[\, \sum^\infty_{a,b=0} Q\Bigl( \Phi^\ast_{(2+n+a+b,\,-1-n-b)} \Phi_{(-1-a-b,\,b)}\Bigr)
\nonumber \\
&+ \Phi^\ast_{(n+1,\,-n)} \bigl(\ez\Phi_{(0,0)}\bigr)
+ \sum^\infty_{k=1} \Phi^\ast_{(n+k+1,\,-n)} \bigl(\ez\Phi_{(-k,0)}\bigr)
\Biggr] + O(g^2)\qquad (n\geq 1)\,,
\end{align}
where the factor $\ez\Xil$ acts as a projector into the small Hilbert space.
Applying this constraint to the fields $\Phi^\ast_{(2+n+b,\,-1-n-b)}$ and $\Phi^\ast_{(n+1,\,-n)}$, 
which are on the right-hand side of \eqref{constraint:n+1,-n},
we have
\bs
\begin{align}
\Phi^\ast_{(2+n+b,\,-1-n-b)} &= O(g)\,,
\label{2+n+b,-1-n-b}
\\[.5ex]
\Phi^\ast_{(n+1,\,-n)} 
&= \delta_{n,1} \Phi^{\ast -}_{(2,-1)} + \Xil \Phi^{\ast\Xi}_{(n+1,\,-n)}
= \delta_{n,1} \Phi^{\ast -}_{(2,-1)} + O(g)\,.
\label{relation:n+1,-n}
\end{align}
\es
Here and in what follows, the symbol $\delta_{n,m}$ denotes the Kronecker delta:
\begin{equation}
\delta_{n,m} := 
\left\{
\begin{aligned}
&1\quad \text{for} \quad n=m\\
&0\quad \text{for} \quad n\neq m
\end{aligned}
\right.
\,.
\end{equation}
(In the first equality of \eqref{relation:n+1,-n}, we have used \eqref{pgf:comp:antighost}.)
Substituting these back into \eqref{constraint:n+1,-n}, we obtain
\begin{align} \label{constraint:n+1,-n:reduc}
\Phi^{\ast\Xi}_{(n+1,-n)}
&= g\,\ez\Xil\Biggl[\, \sum^\infty_{a=1} \sum^\infty_{b=0} Q\Bigl( \Phi^\ast_{(2+n+a+b,\,-1-n-b)} \Phi_{(-1-a-b,\,b)}\Bigr)
\nonumber \\
&+ \delta_{n,1} \Phi^{\ast -}_{(2,-1)} \bigl(\ez\Phi_{(0,0)}\bigr)
+ \sum^\infty_{k=1} \Phi^\ast_{(n+k+1,\,-n)} \bigl(\ez\Phi_{(-k,0)}\bigr)
\Biggr] + O(g^2)\qquad (n\geq 1)\,.
\end{align}
Note that the range of the summation for $a$ got narrower: the index runs from one, not zero.
As can be seen from \eqref{constraint:n+1,-n:reduc}, 
after the Lagrange multiplier $\PhiXi_{(-n,n)}$ is integrated out,
cubic terms including $\Phi^{\ast\Xi}_{(n+1,-n)}$ become $O(g^2)$ and can be neglected in the present analysis. 
Instead, part of the quadratic term $\iu\hs\bigl\langle\Phi^\ast_{(n+1,-n)}, Q\Phi_{(-n,n-1)}\bigr\rangle$ 
in $\calSB_\mrm{quad}$ contributes to terms of order $g$.
Indeed we have
\begin{equation} \label{contribution:n+1,-n}
\iu\,\Bigl\langle\Phi^\ast_{(n+1,-n)}\,,\,Q\Phi_{(-n,n-1)}\Bigr\rangle
= \iu\,\delta_{n,1} \Bigl\langle\Phi^{\ast -}_{(2,-1)}\,,\, Q\Phi_{(-1,0)} \Bigr\rangle
+ \iu\,\Bigl\langle\Xil\Phi^{\ast\Xi}_{(n+1,-n)}\,,\,Q\Phi_{(-n,n-1)}\Bigr\rangle
\end{equation}
with eqs.~\eqref{pgf:comp} and \eqref{constraint:n+1,-n:reduc} imposed,
and the second term on the right-hand side is of order $g$.

Next let us examine the terms including a second Lagrange multiplier field $\Phi_{(-n,n-1)}$ ($n\geq 2$).
Among these terms, those in $\calSB_Q$ do not contribute in the present analysis because
they are of order $g^2$ owing to the constraint \eqref{constraint:n+1,-n:reduc}.
Below is what can be relevant:
\bs \label{terms:-n,n-1}
\begin{align}
& \iu\,\Bigl\langle \Phi^\ast_{(n+1,-(n-1))}, \ez\Phi_{(-n,n-1)}\Bigr\rangle\,,
\label{-n,n-1:quad}
\\[1ex]
& \iu\,g\,\delta_{n,2}\Bigl\langle \Phi^{\ast -}_{(2,-1)} \Phi^{\ast -}_{(2,-1)}\,,\, \Phi_{(-2,1)}\Bigr\rangle\,,
\label{-n,n-1:N}
\\[1ex]
& -\iu\,g \Bigl\langle \Phi^\ast_{(n+1,\,-(n-1))} \bigl(\ez\Phi_{(0,0)}\bigr) \,,\, \Phi_{(-n,n-1)}\Bigr\rangle
- \iu\,g\sum^\infty_{k=1} \Bigl\langle \Phi^\ast_{(n+k+1,\,-(n-1))} \bigl(\ez\Phi_{(-k,0)}\bigr) \,,\, \Phi_{(-n,n-1)}\Bigr\rangle\,,
\label{-n,n-1:eta}
\\[1ex]
& \iu\,g \Biggl\langle\Xil\Biggl[\, \sum^\infty_{a=1} \sum^\infty_{b=0} Q\Bigl( \Phi^\ast_{(2+n+a+b,\,-1-n-b)} \Phi_{(-1-a-b,\,b)}\Bigr)
+ \sum^\infty_{l=1} \Phi^\ast_{(n+l+1,\,-n)} \bigl(\ez\Phi_{(-l,0)}\bigr)
\Biggr]\,,\,Q\Phi_{(-n,n-1)}\Biggr\rangle\,,
\label{-n,n-1:-n,n:quad}
\end{align}
\es
where the terms \eqref{-n,n-1:quad}, \eqref{-n,n-1:N}, \eqref{-n,n-1:eta}, and \eqref{-n,n-1:-n,n:quad}
are the contributions from $\calSB_\mrm{quad}$, $\calSB_N$, $\calSB_\eta$, and \eqref{contribution:n+1,-n}, respectively.
Note that $\calSB_N$ contributes only for $n=2$.
After integrating out $\PhiXi_{(-n,n-1)}$ ($n\geq 2$), we obtain the constraint
\begin{align} \label{constraint:n+1,-(n-1)}
\Phi^{\ast\Xi}_{(n+1, -(n-1))}
=& -g\,\ez\Xil\Biggl[
\delta_{n,2} \Phi^{\ast -}_{(2,-1)} \Phi^{\ast -}_{(2,-1)}
\nonumber \\
&\  
- \Phi^\ast_{(n+1,\,-(n-1))} \bigl(\ez\Phi_{(0,0)}\bigr)
- \sum^\infty_{k=1} \Phi^\ast_{(n+k+1,\,-(n-1))} \bigl(\ez\Phi_{(-k,0)}\bigr)
\nonumber \\
&\ 
+ Q\Xil\biggl\{\, \sum^\infty_{a=1} \sum^\infty_{b=0} Q\Bigl( \Phi^\ast_{(2+n+a+b,\,-1-n-b)} \Phi_{(-1-a-b,\,b)}\Bigr)
+ \sum^\infty_{l=1} \Phi^\ast_{(n+l+1,\,-n)} \bigl(\ez\Phi_{(-l,0)}\bigr)
\biggr\} \Biggr]
\nonumber \\
&+O(g^2) \qquad (n\geq 2)\,.
\end{align}
The crucial point here is that all the terms in \eqref{terms:-n,n-1} are linear in $\Phi_{(-n,n-1)}$,
as was mentioned below \eqref{SB_eta}.
If we did not have \eqref{2+n+b,-1-n-b} and could not neglect the $a=0$ term in \eqref{constraint:n+1,-n},
eq.~\eqref{-n,n-1:-n,n:quad} would include the term
\begin{equation}
\iu\,g \Biggl\langle\Xil Q\Bigl( \Phi^\ast_{(2+n+a+b,\,-1-n-b)} \Phi_{(-1-a-b,\,b)}\Bigr)
\,,\,Q\Phi_{(-n,n-1)}\Biggr\rangle\Biggr|_{a=0,\ b=n-1}\ ,
\end{equation}
which is \emph{quadratic} in $\Phi_{(-n,n-1)}$,
in which case we could not treat $\PhiXi_{(-n,n-1)}$ as a Lagrange multiplier.

Applying the constraint \eqref{constraint:n+1,-(n-1)} 
to the fields $\Phi^\ast_{(n+1,\,-(n-1))}$, $\Phi^\ast_{(3+n+b,\,-1-n-b)}$, and $\Phi^\ast_{(n+2,\,-n)}$,
which are on the right-hand side of \eqref{constraint:n+1,-(n-1)},
we have
\bs
\begin{align}
\Phi^\ast_{(n+1,\,-(n-1))} 
&= \delta_{n,2} \Phi^{\ast -}_{(3,-1)} + \Xil \Phi^{\ast\Xi}_{(n+1,\,-(n-1))} 
= \delta_{n,2} \Phi^{\ast -}_{(3,-1)} +O(g)\,,
\\[1ex]
\Phi^\ast_{(3+n+b,\,-1-n-b)} &= O(g)\,,
\\[1ex]
\Phi^\ast_{(n+2,\,-n)} &= O(g)\,.
\end{align}
\es
Substituting these back into \eqref{constraint:n+1,-(n-1)} and using the relation $\Xil Q\Xil = \Xil\Xl$, we obtain
\begin{align} \label{constraint:n+1,-(n-1):reduc}
\Phi^{\ast\Xi}_{(n+1, -(n-1))}
=& -g\,\ez\Xil\Biggl[
\delta_{n,2} \Bigl( \Phi^{\ast -}_{(2,-1)} \Phi^{\ast -}_{(2,-1)} -  \Phi^{\ast -}_{(3,-1)} \bigl(\ez\Phi_{(0,0)}\bigr) \Bigr)
- \sum^\infty_{k=1} \Phi^\ast_{(n+k+1,\,-(n-1))} \bigl(\ez\Phi_{(-k,0)}\bigr)
\nonumber \\[.5ex]
&\ 
+ \Xl\biggl\{\, \sum^\infty_{a=2} \sum^\infty_{b=0} Q\Bigl( \Phi^\ast_{(2+n+a+b,\,-1-n-b)} \Phi_{(-1-a-b,\,b)}\Bigr)
+ \sum^\infty_{l=2} \Phi^\ast_{(n+l+1,\,-n)} \bigl(\ez\Phi_{(-l,0)}\bigr)
\biggr\} \Biggr]
\nonumber \\[.5ex]
&+O(g^2) \qquad (n\geq 2)\,.
\end{align}
Now the ranges of the summation for $a$ and that for $l$ became narrower.
Because of the relation \eqref{constraint:n+1,-(n-1):reduc},
cubic terms including $\Phi^{\ast\Xi}_{(n+1, -(n-1))}$ become $O(g^2)$ after we integrate out the Lagrange multiplier $\PhiXi_{(-n,n-1)}$, 
and can therefore be neglected in our analysis.
Instead, part of the quadratic term $\iu\hs\bigl\langle\Phi^\ast_{(n+1,-(n-1))}, Q\Phi_{(-n,n-2)}\bigr\rangle$ 
in $\calSB_\mrm{quad}$ contributes to terms of order $g$:
\begin{equation} \label{contribution:n+1,-(n-1)}
\iu\,\Bigl\langle\Phi^\ast_{(n+1,-(n-1))}\,,\,Q\Phi_{(-n,n-2)}\Bigr\rangle
= \iu\,\delta_{n,2} \Bigl\langle\Phi^{\ast -}_{(3,-1)}\,,\, Q\Phi_{(-2,0)} \Bigr\rangle
+ \iu\,\Bigl\langle\Xil\Phi^{\ast\Xi}_{(n+1,-(n-1))}\,,\,Q\Phi_{(-n,n-2)}\Bigr\rangle
\end{equation}
with eqs.~\eqref{pgf:comp} and \eqref{constraint:n+1,-(n-1):reduc} imposed.
It should be added that after \eqref{constraint:n+1,-(n-1):reduc} is taken into account,
eq.~\eqref{constraint:n+1,-n:reduc} becomes
\begin{align} \label{constraint:n+1,-n:reduc:2}
\Phi^{\ast\Xi}_{(n+1,-n)}
&= g\,\ez\Xil\Biggl[\, \sum^\infty_{a=2} \sum^\infty_{b=0} Q\Bigl( \Phi^\ast_{(2+n+a+b,\,-1-n-b)} \Phi_{(-1-a-b,\,b)}\Bigr)
\nonumber \\[.5ex]
&+ \delta_{n,1} \Bigl( \Phi^{\ast -}_{(2,-1)} \bigl(\ez\Phi_{(0,0)}\bigr) + \Phi^{\ast -}_{(3,-1)} \bigl(\ez\Phi_{(-1,0)}\bigr) \Bigr)
\nonumber \\[.5ex]
&+ \sum^\infty_{k=2} \Phi^\ast_{(n+k+1,\,-n)} \bigl(\ez\Phi_{(-k,0)}\bigr)
\Biggr] + O(g^2)\qquad (n\geq 1)\,,
\end{align}
in which the ranges of the summations for $a$ and $k$ got narrower.
To summarize, what we have obtained up to this stage is
\begin{align}
\Phi^{\ast\Xi}_{(n+1,-n)}
&= g\,\ez\Xil\Biggl[\, \sum^\infty_{a=2} \sum^\infty_{b=0} Q\Bigl( \Phi^\ast_{(2+n+a+b,\,-1-n-b)} \Phi_{(-1-a-b,\,b)}\Bigr)
\nonumber \\[.5ex]
&+ \delta_{n,1} \Bigl( \Phi^{\ast -}_{(2,-1)} \bigl(\ez\Phi_{(0,0)}\bigr) + \Phi^{\ast -}_{(3,-1)} \bigl(\ez\Phi_{(-1,0)}\bigr) \Bigr)
\nonumber \\[.5ex]
&+ \sum^\infty_{k=2} \Phi^\ast_{(n+k+1,\,-n)} \bigl(\ez\Phi_{(-k,0)}\bigr)
\Biggr] + O(g^2)\qquad (n\geq 1)\,,
\tag{\ref{constraint:n+1,-n:reduc:2} bis}
\end{align}
\begin{align} 
\Phi^{\ast\Xi}_{(n+1, -(n-1))}
=& -g\,\ez\Xil\Biggl[
\delta_{n,2} \Bigl( \Phi^{\ast -}_{(2,-1)} \Phi^{\ast -}_{(2,-1)} -  \Phi^{\ast -}_{(3,-1)} \bigl(\ez\Phi_{(0,0)}\bigr) \Bigr)
- \sum^\infty_{k=1} \Phi^\ast_{(n+k+1,\,-(n-1))} \bigl(\ez\Phi_{(-k,0)}\bigr)
\nonumber \\[.5ex]
&\ 
+ \Xl\biggl\{\, \sum^\infty_{a=2} \sum^\infty_{b=0} Q\Bigl( \Phi^\ast_{(2+n+a+b,\,-1-n-b)} \Phi_{(-1-a-b,\,b)}\Bigr)
+ \sum^\infty_{l=2} \Phi^\ast_{(n+l+1,\,-n)} \bigl(\ez\Phi_{(-l,0)}\bigr)
\biggr\} \Biggr]
\nonumber \\[.5ex]
&+O(g^2) \qquad (n\geq 2)\,,
\tag{\ref{constraint:n+1,-(n-1):reduc} bis}
\end{align}
\begin{equation} 
\iu\,\Bigl\langle\Phi^\ast_{(n+1,-n)}\,,\,Q\Phi_{(-n,n-1)}\Bigr\rangle
= \iu\,\delta_{n,1} \Bigl\langle\Phi^{\ast -}_{(2,-1)}\,,\, Q\Phi_{(-1,0)} \Bigr\rangle
+ \iu\,\Bigl\langle\Xil\Phi^{\ast\Xi}_{(n+1,-n)}\,,\,Q\Phi_{(-n,n-1)}\Bigr\rangle
\tag{\ref{contribution:n+1,-n} bis}
\end{equation}
with eqs.~\eqref{pgf:comp} and \eqref{constraint:n+1,-n:reduc:2} imposed,
and
\begin{equation} 
\iu\,\Bigl\langle\Phi^\ast_{(n+1,-(n-1))}\,,\,Q\Phi_{(-n,n-2)}\Bigr\rangle
= \iu\,\delta_{n,2} \Bigl\langle\Phi^{\ast -}_{(3,-1)}\,,\, Q\Phi_{(-2,0)} \Bigr\rangle
+ \iu\,\Bigl\langle\Xil\Phi^{\ast\Xi}_{(n+1,-(n-1))}\,,\,Q\Phi_{(-n,n-2)}\Bigr\rangle
\tag{\ref{contribution:n+1,-(n-1)} bis}
\end{equation}
with eqs.~\eqref{pgf:comp} and \eqref{constraint:n+1,-(n-1):reduc} imposed.

Let us go one more step further. We examine the terms including a third Lagrange multiplier field $\Phi_{(-n,n-2)}$ ($n\geq 3$).
Owing to \eqref{constraint:n+1,-n:reduc:2} and \eqref{constraint:n+1,-(n-1):reduc},
those in $\calSB_Q$ do not contribute to terms of order $g$, and $\calSB_N$ contributes only for $n=3$.
Therefore, what can be relevant is
\bs
\begin{align}
& \iu\,\Bigl\langle \Phi^\ast_{(n+1,-(n-2))}, \ez\Phi_{(-n,n-2)}\Bigr\rangle\,,
\label{-n,n-2:quad}
\\[1ex]
& \iu\,g\,\delta_{n,3}\Bigl\langle \bigl\{\Phi^{\ast -}_{(2,-1)}, \Phi^{\ast -}_{(3,-1)}\bigr\}\,,\, \Phi_{(-3,1)}\Bigr\rangle\,,
\label{-n,n-2:N}
\\[1ex]
& -\iu\,g \Bigl\langle \Phi^\ast_{(n+1,\,-(n-2))} \bigl(\ez\Phi_{(0,0)}\bigr) \,,\, \Phi_{(-n,n-2)}\Bigr\rangle
- \iu\,g\sum^\infty_{k=1} \Bigl\langle \Phi^\ast_{(n+k+1,\,-(n-2))} \bigl(\ez\Phi_{(-k,0)}\bigr) \,,\, \Phi_{(-n,n-2)}\Bigr\rangle\,,
\label{-n,n-2:eta}
\\[1ex]
& \iu\,g \Biggl\langle\Xil\Biggl[\,
\sum^\infty_{k=1} \Phi^\ast_{(n+k+1,\,-(n-1))} \bigl(\ez\Phi_{(-k,0)}\bigr)
\nonumber \\[.5ex]
&\ 
- \Xl\biggl\{\, \sum^\infty_{a=2} \sum^\infty_{b=0} Q\Bigl( \Phi^\ast_{(2+n+a+b,\,-1-n-b)} \Phi_{(-1-a-b,\,b)}\Bigr)
+ \sum^\infty_{l=2} \Phi^\ast_{(n+l+1,\,-n)} \bigl(\ez\Phi_{(-l,0)}\bigr)
\biggr\} \Biggr]\,,\,Q\Phi_{(-n,n-2)}\Biggr\rangle\,,
\label{-n,n-2:-n,n-1:quad}
\end{align}
\es
where the terms \eqref{-n,n-2:quad}, \eqref{-n,n-2:N}, \eqref{-n,n-2:eta}, and \eqref{-n,n-2:-n,n-1:quad}
are the contributions from $\calSB_\mrm{quad}$, $\calSB_N$, $\calSB_\eta$, and \eqref{contribution:n+1,-(n-1)}, respectively.
Integrating out $\PhiXi_{(-n,n-2)}$ ($n\geq 3$), we obtain the constraint
\begin{align} \label{constraint:n+1,-(n-2)}
\Phi^{\ast\Xi}_{(n+1,-(n-2))}
=&  -g\,\ez\Xil\Biggl[
\delta_{n,3} \bigl\{\Phi^{\ast -}_{(2,-1)}, \Phi^{\ast -}_{(3,-1)}\bigr\}
\nonumber \\
&\  
- \Phi^\ast_{(n+1,\,-(n-2))} \bigl(\ez\Phi_{(0,0)}\bigr)
- \sum^\infty_{j=1} \Phi^\ast_{(n+j+1,\,-(n-2))} \bigl(\ez\Phi_{(-j,0)}\bigr)
\nonumber \\
&\ 
+ Q\Xil\Biggl(\, 
\sum^\infty_{k=1} \Phi^\ast_{(n+k+1,\,-(n-1))} \bigl(\ez\Phi_{(-k,0)}\bigr)
\nonumber \\[.5ex]
&\ 
- \Xl\biggl\{ \sum^\infty_{a=2} \sum^\infty_{b=0} Q\Bigl( \Phi^\ast_{(2+n+a+b,\,-1-n-b)} \Phi_{(-1-a-b,\,b)}\Bigr)
+ \sum^\infty_{l=2} \Phi^\ast_{(n+l+1,\,-n)} \bigl(\ez\Phi_{(-l,0)}\bigr)
\biggr\}
\Biggr) \Biggr]
\nonumber \\
&+O(g^2) 
\nonumber \\
=&  -g\,\ez\Xil\Biggl[
\delta_{n,3} \Bigl(\bigl\{\Phi^{\ast -}_{(2,-1)}, \Phi^{\ast -}_{(3,-1)}\bigr\}
- \Phi^\ast_{(4,-1)} \bigl(\ez\Phi_{(0,0)}\bigr)\Bigr)
\nonumber \\
&\  
- \sum^\infty_{j=1} \Phi^\ast_{(n+j+1,\,-(n-2))} \bigl(\ez\Phi_{(-j,0)}\bigr)
\nonumber \\
&\ 
+ \Xl\Biggl(\, 
\sum^\infty_{k=2} \Phi^\ast_{(n+k+1,\,-(n-1))} \bigl(\ez\Phi_{(-k,0)}\bigr)
\nonumber \\[.5ex]
&\ 
- \Xl\biggl\{ \sum^\infty_{a=3} \sum^\infty_{b=0} Q\Bigl( \Phi^\ast_{(2+n+a+b,\,-1-n-b)} \Phi_{(-1-a-b,\,b)}\Bigr)
+ \sum^\infty_{l=3} \Phi^\ast_{(n+l+1,\,-n)} \bigl(\ez\Phi_{(-l,0)}\bigr)
\biggr\}
\Biggr) \Biggr]
\nonumber \\
&+O(g^2) 
\qquad (n\geq 3)\,.
\end{align}
In the second equality, we performed the deformation similar to that we did 
to obtain \eqref{constraint:n+1,-(n-1):reduc} from \eqref{constraint:n+1,-(n-1)}.
Note the ranges of the summations for $a$, $k$, and $l$.
The constraint \eqref{constraint:n+1,-(n-2)} tells us that
after the Lagrange multiplier $\PhiXi_{(-n,n-2)}$ is integrated out,
cubic terms including $\Phi^{\ast\Xi}_{(n+1,-(n-2))}$ become $O(g^2)$ and can be neglected, 
but the quadratic term $\iu\,\bigl\langle\Phi^\ast_{(n+1,-(n-2))}\,,\,Q\Phi_{(-n,n-3)}\bigr\rangle$ in $\calSB_\mrm{quad}$
includes an $O(g)$ part:
\begin{equation} 
\iu\,\Bigl\langle\Phi^\ast_{(n+1,-(n-2))}\,,\,Q\Phi_{(-n,n-3)}\Bigr\rangle
= \iu\,\delta_{n,3} \Bigl\langle\Phi^{\ast -}_{(4,-1)}\,,\, Q\Phi_{(-3,0)} \Bigr\rangle
+ \iu\,\Bigl\langle\Xil\Phi^{\ast\Xi}_{(n+1,-(n-2))}\,,\,Q\Phi_{(-n,n-3)}\Bigr\rangle
\end{equation}
with \eqref{pgf:comp} and \eqref{constraint:n+1,-(n-2)} imposed.
Furthermore, if we take account of \eqref{constraint:n+1,-(n-2)},
eqs.~\eqref{constraint:n+1,-n:reduc:2} and \eqref{constraint:n+1,-(n-1):reduc} become as follows:
\begin{align}
\Phi^{\ast\Xi}_{(n+1,-n)}
&= g\,\ez\Xil\Biggl[ \delta_{n,1} \Bigl( \Phi^{\ast -}_{(2,-1)} \bigl(\ez\Phi_{(0,0)}\bigr) 
+ \Phi^{\ast -}_{(3,-1)} \bigl(\ez\Phi_{(-1,0)}\bigr) 
+ \Phi^{\ast -}_{(4,-1)} \bigl(\ez\Phi_{(-2,0)}\bigr) \Bigr)
\nonumber \\[.5ex]
&+ \sum^\infty_{a=3} \sum^\infty_{b=0} Q\Bigl( \Phi^\ast_{(2+n+a+b,\,-1-n-b)} \Phi_{(-1-a-b,\,b)}\Bigr)
\nonumber \\[.5ex]
&+ \sum^\infty_{k=3} \Phi^\ast_{(n+k+1,\,-n)} \bigl(\ez\Phi_{(-k,0)}\bigr)
\Biggr] + O(g^2)\qquad (n\geq 1)\,,
\label{constraint:n+1,-n:reduc:3} \\[1ex]
%---
\Phi^{\ast\Xi}_{(n+1, -(n-1))}
&= g\,\ez\Xil\Biggl[
\delta_{n,2} \Bigl( -\Phi^{\ast -}_{(2,-1)} \Phi^{\ast -}_{(2,-1)} 
+ \Phi^{\ast -}_{(3,-1)} \bigl(\ez\Phi_{(0,0)}\bigr) 
+ \Phi^{\ast -}_{(4,-1)} \bigl(\ez\Phi_{(-1,0)}\bigr) \Bigr)
\nonumber \\[.5ex]
&+ \sum^\infty_{k=2} \Phi^\ast_{(n+k+1,\,-(n-1))} \bigl(\ez\Phi_{(-k,0)}\bigr)
\nonumber \\[.5ex]
&\ 
- \Xl\biggl\{\, \sum^\infty_{a=3} \sum^\infty_{b=0} Q\Bigl( \Phi^\ast_{(2+n+a+b,\,-1-n-b)} \Phi_{(-1-a-b,\,b)}\Bigr)
+ \sum^\infty_{l=3} \Phi^\ast_{(n+l+1,\,-n)} \bigl(\ez\Phi_{(-l,0)}\bigr)
\biggr\} \Biggr]
\nonumber \\[.5ex]
&+O(g^2) \qquad (n\geq 2)\,.
\label{constraint:n+1,-(n-1):reduc:2}
\end{align}
In this manner, 
as we integrate out Lagrange multipliers,
cubic terms including $\Phi^{\ast \Xi}_{(n+1, -m)}$ ($1\leq m\leq n$) drop away,
and $\calSB_\mrm{quad}$ generates $O(g)$ terms;
moreover, the ranges of the infinite series in the constraints become narrower step by step.
The ultimate forms of the constraints are given by 
\begin{align} \label{constraint:final}
\Phi^{\ast\Xi}_{(n+1, -m)}
&= g\,\delta_{m,1}\,\ez\Xil\Biggl[ -\sum^n_{p=2} \Phi^{\ast -}_{(p,-1)} \Phi^{\ast -}_{(n-p+2,-1)}
+ \sum^\infty_{q=0} \Phi^{\ast -}_{(n+1+q, -1)} \bigl( \ez\Phi_{(-q,0)}\bigr) \Biggl]
\nonumber \\
& +O(g^2) 
\qquad (1\leq m\leq n)\,,
\end{align}
with
\begin{equation}
\sum^1_{p=2}( \cdots ) := 0\,.
\end{equation}
Thus, among the fields $\Phi^{\ast\Xi}_{(n+1, -m)}$, only $\Phi^{\ast\Xi}_{(n+1, -1)}$ are relevant in our analysis.
We finally find that the completely reduced action $\calSB_\mrm{red}$ takes the form
\begin{align} \label{SB_red}
\calSB_\mrm{red}
&= -\frac{\iu}{2}\, \bigl\langle \Phi_{(0,0)}, Q\ez\Phi_{(0,0)}\bigr\rangle
+\frac{\iu}{6}\,g\,\Bigl\langle \ez\Phi_{(0,0)}\,,\, \bigl[ \Phi_{(0,0)}, Q\Phi_{(0,0)}\bigr]\Bigr\rangle
\nonumber \\[.5ex]
&+\iu\,\sum^\infty_{n=1}\Bigl\langle\Phi^\ast_{(n+1,-1)}\,,\,Q\Phi_{(-n,0)}\Bigr\rangle +O(g^2)
\end{align}
with the conditions \eqref{pgf:comp} and the constraints \eqref{constraint:final} imposed.
The first two terms on the right-hand side are nothing but the quadratic term and the cubic term in the original action $\SB$.
The relation between the original actions $\SB$ and $\SW$ has already been manifested in the previous paper~\cite{INOT}:
under the partial gauge fixing \eqref{cond:0,0}, the action $\SB$ can be regarded as the regularized version of $\SW$,
and we have
\begin{equation}
-\frac{\iu}{2}\, \bigl\langle \Phi_{(0,0)}, Q\ez\Phi_{(0,0)}\bigr\rangle
+\frac{\iu}{6}\,g\,\Bigl\langle \ez\Phi_{(0,0)}\,,\, \bigl[ \Phi_{(0,0)}, Q\Phi_{(0,0)}\bigr]\Bigr\rangle
\, \longrightarrow \, \SW
\qquad (\lambda \to 0)\,.
\end{equation}
Therefore, in what follows, we concentrate on the third term on the right-hand side of \eqref{SB_red}.
It can be written as
\begin{equation} \label{SB_red:3rd}
\iu\,\sum^\infty_{n=1}\Bigl\langle\Phi^\ast_{(n+1,-1)}\,,\,Q\Phi_{(-n,0)}\Bigr\rangle
= \iu\,\sum^\infty_{n=1}\Bigl\langle\Phi^{\ast -}_{(n+1,-1)}\,,\,Q\Xil\PhiXi_{(-n,0)}\Bigr\rangle
+ \iu\,\sum^\infty_{n=1}\Bigl\langle\Xil \Phi^{\ast\Xi}_{(n+1,-1)}\,,\,Q\Xil\PhiXi_{(-n,0)}\Bigr\rangle\,.
\end{equation}
After the same calculation as performed in \eqref{reduced:SB_1},
the first term on the right-hand side reduces to the following form:
\begin{equation} \label{SB_red:quadratic}
\iu\,\sum^\infty_{n=1}\Bigl\langle\Phi^{\ast -}_{(n+1,-1)}\,,\,Q\Xil\PhiXi_{(-n,0)}\Bigr\rangle
= \sum^\infty_{n=1}\Bllangle\Phi^{\ast -}_{(n+1,-1)} \,,\, Q\PhiXi_{(-n,0)}\Brrangle
= - \sum^\infty_{n=1}\Bllangle\psi_{n+1} \,,\, Q\psi_{-n+1}\Brrangle\,,
\end{equation}
where we have defined $\psi_n$ ($n\in \mathbb{Z}$) by
\begin{equation}
\Phi^{\ast -}_{(n+1,-1)} =-\psi_{n+1}\,,\quad
\PhiXi_{(-n,0)} = \psi_{-n+1}
\qquad (n\geq 0)\,.
\end{equation}
Eq.~\eqref{SB_red:quadratic} is equal to the sum of the quadratic terms \eqref{SW_n}
under the identification
\begin{equation} \label{psi:Psi}
\psi_n \cong \PsiW_{(n,-1)}\quad (\forall n\in \mathbb{Z})\,,
\end{equation}
which is equivalent to \eqref{identification}.
Furthermore, the second term on the right-hand side of \eqref{SB_red:3rd} can be deformed as
\begin{align} \label{SB_red:cubic}
&\quad \iu\,\sum^\infty_{n=1}\,\Bigl\langle\Xil \Phi^{\ast\Xi}_{(n+1,-1)}\,,\,Q\Xil\PhiXi_{(-n,0)}\Bigr\rangle
= \iu\,\sum^\infty_{n=1}\,\Bigl\langle\Xil \Phi^{\ast\Xi}_{(n+1,-1)}\,,\,\Xl\PhiXi_{(-n,0)}\Bigr\rangle
\nonumber \\[.5ex]
=& -\iu\,g\sum^\infty_{n=1}\,\biggl\langle\Xil\biggl[\,
\sum^n_{p=2} \psi_p \psi_{n-p+2} + \sum^\infty_{q=0}\psi_{n+q+1} \psi_{-q+1}\biggr]\,,\, \Xl\psi_{-n+1}\biggr\rangle
+O(g^2)
\nonumber \\[.5ex]
=& -g\sum^\infty_{n=1}\,\bgllangle \sum^n_{p=2} \psi_p \psi_{n-p+2} + \sum^\infty_{q=0}\psi_{n+q+1} \psi_{-q+1}\,,\, \Xl\psi_{-n+1}\bgrrangle
+O(g^2)\,.
\end{align}
In the second equality, we have used \eqref{constraint:final}.
Rewriting the cubic term in \eqref{Witten:master action} as
\begin{align} \label{Witten:master action:cubic}
&-\frac{g}{3} \Bllangle \bPsiW, \Xmid\bigl(\bPsiW\ast\bPsiW\bigr) \Brrangle
\nonumber \\[1ex]
=& -\frac{g}{3} \Bllangle \PsiW_{(1,-1)}, \Xmid\bigl(\PsiW_{(1,-1)}\ast\PsiW_{(1,-1)}\bigr) \Brrangle
\nonumber \\[.5ex]
&-g\sum^\infty_{n=1}\,\bgllangle \sum^n_{p=2} \PsiW_{(p,-1)} \PsiW_{(n-p+2,-1)} 
+ \sum^\infty_{q=0}\PsiW_{(n+q+1,-1)} \PsiW_{(-q+1,-1)}\,,\, \Xmid\,\PsiW_{(-n+1,-1)}\bgrrangle\,,
\end{align}
we find that in the singular limit $\lambda\to 0$, 
eq.~\eqref{SB_red:cubic} coincides with the cubic term in the master action $\calSW$ minus the one in the original action $\SW$
up to $O(g^2)$ terms under the identification \eqref{psi:Psi}, namely \eqref{identification}.
We thus have 
\begin{equation}
\lim_{\lambda\to 0}\calSB = \calSW + O(g^2) \,.
\end{equation}
Because the Berkovits formulation is regular regardless of the presence of the interaction, 
from the above result we can conclude that $\calSB$ is the regularized version of $\calSW$,
with the terms higher order in $g$ in $\calSB$ playing the role of the counterterms for canceling the divergences in the Witten formulation.

%%%%%%%%%%%%%%%%%%%%%%%%%%%%%%%%%%%%%%%%%
\section{Summary and discussion}
\label{section:discussion}
\setcounter{equation}{0}
For the purpose of acquiring a deeper understanding of 
the relation between the small Hilbert space approach and the large Hilbert space approach to open superstring field theory,
in the present paper we have investigated the Berkovits formulation in detail with the technique of partial gauge fixing,
and have clarified its relation to the Witten formulation at the level of the reducibility structure and the master action.
In particular, the master action in the Berkovits formulation has turned out to reduce to
the regularized version of that in the Witten formulation after partial gauge fixing. 
As shown in subsection~\ref{relation:reducibility}, behind this relation is the correspondence between
a reducibility sub-structure of the Berkovits formulation and the reducibility structure of the Witten formulation.
In general, the form of a reducibility structure governs that of a master action.

For our analysis of the master action in the Berkovits formulation, 
it was sufficient to investigate its quadratic terms and cubic terms.
In fact, its higher-order terms have not been completely obtained yet.\footnote{
For some approaches to this problem, see refs.~\cite{paperII, constrainedBV}.
}
We expect, however, that our result will be useful also for solving this problem.
In order to see the point in which the difficulty lies, 
let us review the way to construct the master action in general.
In principle, one can construct a master action $\calS$, namely a solution to the classical master equation 
in the BV formalism systematically in the following manner.
First, one expands $\calS$ in what is called the antifield number:
\begin{equation}
\calS = \sum^\infty_{n=0} \calS^{(n)}\,,\quad \calS^{(0)} = S\,,
\end{equation}
where $\calS^{(n)}$ $(n\geq 0)$ denotes the sum of all the terms of antifield number $n$, with
$\calS^{(0)}$ coinciding with the original action $S$.
Consequently, the master equation is decomposed into its subequations.
Then, using the subequations, one can determine $\calS^{(N+1)}$ $(N\geq 0)$ if one knows 
the form of the reducibility structure and the actions $\calS^{(0)}$, $\calS^{(1)}$, ..., $\calS^{(N)}$.
Thus, given an $S$ $(= \calS^{(0)})$ and a reducibility structure, all the $\calS^{(n)}$'s can be obtained one by one.
In some theories such as Yang-Mills theories, only a finite number of $\calS^{(n)}$'s are nonzero:
\begin{equation} \label{S^n}
\calS^{(n)} = 0 \quad \left( \forall n \geq \exists n_0 \geq 0\right).
\end{equation}
In this case, one can solve the master equation completely, merely by carrying out the above-mentioned procedure.
There are, however, other theories in which \eqref{S^n} does not hold:
no matter how large $n_0 \geq 0$ one may take, there exists some $n$ $(\geq n_0)$ satisfying $\calS^{(n)}\neq 0$,
and hence the procedure cannot terminate at any finite step.
In fact, string field theory is exactly one of such complicated gauge theories.
A strategy to find a solution in this class of theories is as follows.
\begin{enumerate}
\item First, one computes $\calS^{(0)}$, $\calS^{(1)}$, ..., and $\calS^{(N)}$ for some $N\geq 0$.
\item Second, from the form of $\sum^N_{n=0} \calS^{(n)}$, one infers the complete solution $\calS$.
\item Third, one confirms that the inferred $\calS$ satisfies the master equation.
\end{enumerate}
In open bosonic string field theory~\cite{Witten:bosonic}, 
the second step is readily performed, and the above strategy works successfully.
In fact, the solution $\calS$ is of the same form as the original action $S$:
one can obtain $\calS$ merely by eliminating the world-sheet ghost number constraint on the string field 
in the original action~\cite{Baulieu, Bochicchio1, Bochicchio2, Thorn:GF, BS}.
It should be noted that behind this result is the mathematical structure 
called $A_\infty$~\cite{Stasheff:I, Stasheff:II, G-J, Markl, P-S, Gaberdiel:1997ia}.\footnote{
Closed bosonic string field theory has another mathematical structure, $L_\infty$,
and its master action can also be readily obtained in virtue of this structure~\cite{Lada:1992wc, S-S, Zwiebach:closed}.}
In the Berkovits formulation of open superstring field theory, however, this kind of structure is obscure;\footnote{
Recently, Erler, Konopka, and Sachs have formulated a new open superstring field theory 
which possesses a manifest $A_\infty$ structure~\cite{EKS}.
In this theory, the master action can be constructed trivially,
but is difficult to treat because of the existence of complicated non-associative multi-string products.
By contrast, the master action in the Berkovits formulation, if it is constructed, must be easy to treat,
formulated only in terms of Witten's star product.
}
as shown in refs.~\cite{paperI, Torii:validity, paperII, Torii:proceedings}, the solution cannot take the same form as the original action,
and it is difficult to infer the complete solution.
In this situation, crucial it can be \emph{how} one performs the first step of the above strategy.
One may think that whatever calculation procedure one may adopt, the resultant $\calS^{(n)}$'s are the same,
and that the more $\calS^{(n)}$'s one obtains, the easier it will become to infer the complete solution.
The fact is, however, that the result does depend on the manner in which one performs the calculation: 
There are degrees of freedom of canonical transformations in the BV formalism, 
which are essentially the degrees of freedom of field redefinition,
and therefore the solution to the master equation is \emph{not} unique.
Thus, what form of a solution one obtains depends on how one calculates.
In particular, what expression of the reducibility structure one starts from is very important.
The point is that only certain forms of partial solutions may be appropriate to infer the complete solution.
We expect, however, our result in the present paper will be useful for approaching this problem.
Starting from the reducibility structure of the expression \eqref{deln:Berkovits:full}
and adopting the calculation procedure which keeps the relation to the Witten formulation manifest,
we will be able to fix most of the degrees of freedom of canonical transformations.

%%%%%%%%%%%%%%%%%%%%%%%%%%%%%%%%%%%%%%%%%
\section*{Acknowledgments}
The work of S.~T.\ was supported in part by the Special Postdoctoral Researcher Program at RIKEN.

%%%%%%%%%%%%%%%%%%%%%%%%%%%%%%%%%%%%%%%%%
\appendix
%%%%%%%%%%%%%%%%%%%%%%%%%%%%%%%%%%%%%%%%%
\section{Properness of the condition \eqref{XiPhi=0} for partial gauge fixing}
\label{properness:pgf}
\setcounter{equation}{0}
In the present appendix, we prove the properness of the condition \eqref{XiPhi=0} for partial gauge fixing: 
we show that for any string field $\Phi = \Phi_{(0,0)}$, there exists a gauge transform $\Phi + \delta\Phi$
which satisfies
\begin{equation} \label{prop:pgf:app}
\Xi\bigl( \Phi + \delta\Phi\bigr)=0\,.
\end{equation}
For this purpose, consider the gauge transformation \eqref{dP} with $\epsilon_{(-1,0)}$ zero:
\begin{equation} \label{dP:app} 
\delta \Phi = \frac{\adg}{1-\e^{-\adg}}\, \ez\epsilon_{(-1,1)} \,.
\end{equation}
Then \eqref{prop:pgf:app} becomes
\begin{equation} \label{eq:ez:-1,1}
\Xi\biggl[ \Phi + \frac{\adg}{1-\e^{-\adg}}\, \ez\epsilon_{(-1,1)} \biggr] =0\,,
\end{equation}
which can be regarded as an equation for $\ez\hs\epsilon_{(-1,1)}$.
For showing the properness, it is sufficient to prove the existence of a solution $\ez\hs\epsilon_{(-1,1)}$ to \eqref{eq:ez:-1,1}.
In fact, we can explicitly construct the solution in the following manner:
First, note that \eqref{eq:ez:-1,1} is equivalent to
\begin{equation}
\ez\epsilon_{(-1,1)} = \ez\Xi \biggl[ \ez\epsilon_{(-1,1)} - \biggl(\Phi + \frac{\adg}{1-\e^{-\adg}}\, \ez\epsilon_{(-1,1)}\biggr) \biggr] =0\,,
\end{equation}
that is,
\begin{equation} \label{rec:ez:-1,1}
\ez\epsilon_{(-1,1)} = -\ez\Xi \biggl[  \Phi + \biggr(\frac{\adg}{1-\e^{-\adg}} -1\biggr) \ez\epsilon_{(-1,1)} \biggr] =0\,.
\end{equation}
Then, we can solve \eqref{rec:ez:-1,1} recursively as
\begin{align} \label{recursion:appA}
\ez\epsilon_{(-1,1)} &= -\sum^\infty_{n=0}\biggl[ -\ez\Xi\biggl(\frac{-\adg}{\e^{-\adg}-1} -1\biggr)\biggr]^n
\ez\Xi\hs\Phi
\nonumber \\[.5ex]
&= - \biggl[ 1 + \ez\Xi\biggl(\frac{-\adg}{\e^{-\adg}-1} -1\biggr)\biggr]^{-1}
\ez\Xi\hs\Phi\,.
\end{align}
In the last equality, we have used the fact that 
the sum in \eqref{recursion:appA} converges for small $g$
because the operator 
\begin{equation}
-\ez\Xi\biggl(\frac{-\adg}{\e^{-\adg}-1} -1\biggr) 
\end{equation}
is $O(g)$.
Thus we have completed the proof of the properness.

%%%%%%%%%%%%%%%%%%%%%%%%%%%%%%%%%%%%%%%%%
\section{Proof of the properness of condition \eqref{Xil:LQ}}
\label{properness}
\setcounter{equation}{0}
Because the gauge parameter $\epsilon_{(-1,0)}$ in the Berkovits formulation resides in the large Hilbert space,
it can be decomposed as 
\begin{equation}
\epsilon_{(-1,0)} = \bigl\{ \ez, \Xil\bigr\} \epsilon_{(-1,0)}
= \ez\Xil \epsilon_{(-1,0)} + \Xil\bigl(\ez\epsilon_{(-1,0)}\bigr)\,,
\end{equation}
and thus it contains two components: $\ez\Xil \epsilon_{(-1,0)}$ in $\Hs$ and $\Xil\bigl(\ez\epsilon_{(-1,0)}\bigr)$ in $\Xil\Hs$.
Nevertheless, all the possible gauge transformations of the form \eqref{gauge transf}
can be covered only by the latter component of $\epsilon_{(-1,0)}$ and the other gauge parameter $\epsilon_{(-1,1)}$,
which moves around the whole of the large Hilbert space.
Hence, as far as the gauge transformation is concerned, without loss of generality
we may assume $\epsilon_{(-1,0)}\in\Xil\Hs$, that is, 
\begin{equation} \label{Xil:LQ:app}
\Xil\epsilon_{(-1,0)} = 0\,.
\end{equation}
In the present appendix, we prove the above claim.
We will show that if we take into account what is called trivial gauge transformations, we can indeed impose
condition \eqref{Xil:LQ:app}, namely \eqref{Xil:LQ}.

%------------------------------------------------------
\subsection{Trivial gauge transformations}
\label{concept:trv}
Before proving the properness of \eqref{Xil:LQ:app}, in the present subsection we will explain
the concept of trivial gauge transformations because it plays a crucial role in our proof. 
In order to explain the concept,
let us consider a system described by a classical action $S = S[\varphi]$, 
which depends on classical bosonic fields $\varphi^i$, with the index $i$ distinguishing different fields.\footnote{
Just for simplicity, we consider only the case in which all the fields $\varphi^i$ are bosonic.
}
The action $S$ is invariant under transformations of the form
\begin{equation}
\varphi^i\ \longrightarrow \ \varphi^i + \delta\varphi^i
\end{equation}
if and only if we have
\begin{equation} \label{deltaS}
\delta S = \sum_i\int \frac{\delta S}{\delta\varphi^i}\hs\delta\varphi^i = 0\,.
\end{equation}
Usually, among transformations satisfying \eqref{deltaS},
only gauge transformations are considered.
There is, however, another type of transformation which is not classified as gauge transformation in the usual sense.
Indeed, we can readily find that \eqref{deltaS} is trivially satisfied by variations of the form
\begin{equation} \label{trv}
\delta_\mrm{trv} \varphi^i = \sum_j A^{ij}\frac{\delta S}{\delta \varphi^j}
\qquad \text{with}\qquad A^{ji} = -A^{ij}\,.
\end{equation}
Because these variations vanish under the use of the equations of motion
\begin{equation}
\frac{\delta S}{\delta\varphi^i} = 0\,,
\end{equation}
they do not entail any Noether identities, and thus they are not genuine gauge transformations.
Nevertheless, the transformations of this kind is called trivial ``gauge'' transformations by convention.
The characteristic of these transformations is that the corresponding variations $\delta_\mrm{trv}\varphi^i$
become identically zero when we use the equations of motion. 
We would like to remark that the degrees of freedom of trivial gauge transformations exist even in non-gauge theories.

%-------------------------------------------------------
\subsection{Proof of the properness}
\label{proof}
In the preceding subsection, we have introduced trivial gauge transformations.
We usually do not consider such transformations because they are not related to genuine gauge degrees of freedom.
In our proof of the properness of \eqref{Xil:LQ:app}, however, they play a crucial role:
it is in virtue of the degrees of freedom of trivial gauge transformations 
that we can impose condition \eqref{Xil:LQ:app} on the parameter $\epsilon_{(-1,0)}$ of the genuine gauge transformation \eqref{gauge transf}
without loss of generality.
The point is that two genuine gauge transformations are equivalent if their difference is merely a trivial gauge transformation.
Our proof is composed of two parts:
\begin{enumerate}
\item
First, we show that for any $\epsilon_{(-1,0)}$, there exists a transform 
$\epsilon_{(-1,0)} + \delta_1\epsilon_{(-1,0)}$ by \eqref{del1:Berkovits:full}
which satisfies 
\begin{equation} \label{Xil:transf}
\Xil\Bigl( \epsilon_{(-1,0)} + \delta_1\epsilon_{(-1,0)}\Bigr) = 0\,.
\end{equation}
\item
Then we show that the gauge transformation specified by the parameters $\epsilon_{(-1,0)}$ and $\epsilon_{(-1,1)}$
and the one specified by their transforms $\epsilon_{(-1,0)} + \delta_1\epsilon_{(-1,0)}$ and $\epsilon_{(-1,1)} + \delta_1\epsilon_{(-1,1)}$
differ merely by a trivial gauge transformation, and therefore the two gauge transformations are equivalent. 
\end{enumerate}
The second part ensures that all the genuine gauge transformations can be generated by 
$\epsilon_{(-1,0)} + \delta_1\epsilon_{(-1,0)}$ and $\epsilon_{(-1,1)} + \delta_1\epsilon_{(-1,1)}$.
Therefore, together with the first part, we can conclude that all the possible gauge transformations of the form \eqref{gauge transf}
can be covered by the $\epsilon_{(-1,0)}$ satisfying \eqref{Xil:LQ:app} and the other parameter $\epsilon_{(-1,1)}$.
It should be noted that in the following analysis, the string field $\Phi$ is not subject to any constraints, 
such as the condition \eqref{gfc} for partial gauge fixing.

In order to prove the first part, 
we consider transformation \eqref{del1:Berkovits:full} with $\epsilon_{(-2,0)}$ and $\epsilon_{(-2,2)}$ zero:
\bs \label{del1:app}
\begin{align} 
\delta_1 \epsilon_{(-1,0)} &= \e^{\adg}\ez\epsilon_{(-2,1)}\,,
\label{del1:LQ} \\
\delta_1 \epsilon_{(-1,1)} &= \tQ\epsilon_{(-2,1)}\qquad \bigl( \tQ := \e^{-\adg}Q\e^{\adg}\bigr).
\end{align}
\es
What we have to do is to show the existence of an $\ez\epsilon_{(-2,1)}$ 
such that the transform $\epsilon_{(-1,0)} + \delta_1\epsilon_{(-1,0)}$ by \eqref{del1:app}
satisfies \eqref{Xil:transf}:
\begin{equation} \label{Xil:transf:2}
\Xil\Bigl( \epsilon_{(-1,0)} + \e^{\adg}\ez\epsilon_{(-2,1)}\Bigr) 
= 0\,.
\end{equation}
In other words, the problem is to show that eq.~\eqref{Xil:transf:2}, which can be regarded as an equation for $\ez\epsilon_{(-2,1)}$,
has a solution. 
In fact, we can explicitly construct a solution $\ez\epsilon_{(-2,1)}$ to \eqref{Xil:transf:2} in the following manner:
First, note that \eqref{Xil:transf:2} is equivalent to
\begin{equation} 
\ez\epsilon_{(-2,1)}
= \ez\Xil\Bigl[ \ez\epsilon_{(-2,1)} - \Bigl( \epsilon_{(-1,0)} + \e^{\adg}\ez\epsilon_{(-2,1)} \Bigr)\Bigr]\,,
\end{equation}
that is, 
\begin{equation} \label{eq:app}
\ez\epsilon_{(-2,1)}
= -\ez\Xil\Bigl[\epsilon_{(-1,0)} + \bigl(\e^{\adg} -1 \bigr)\ez\epsilon_{(-2,1)}\Bigr]\,.
\end{equation}
Then, we can solve \eqref{eq:app} recursively as
\begin{align} \label{recursion:app}
\ez\epsilon_{(-2,1)} 
&= -\sum^\infty_{n=0} \Bigl[-\ez\Xil\bigl(\e^{\adg}-1\bigr)\Bigr]^n\ez\Xil \epsilon_{(-1,0)} 
= -\Bigl[1+ \ez\Xil\bigl(\e^{\adg}-1\bigr)\Bigr]^{-1} \ez\Xil \epsilon_{(-1,0)}\,.
\end{align}
The explicit form of the transform $\epsilon_{(-1,0)} + \delta_1\epsilon_{(-1,0)}$ by \eqref{del1:app} with \eqref{recursion:app}
is given by
\begin{align}
\epsilon_{(-1,0)} + \delta_1\epsilon_{(-1,0)}
&= \bigl(\Xil\ez + \ez\Xil\bigr) \epsilon_{(-1,0)} - \e^{\adg}\Bigl[1+ \ez\Xil\bigl(\e^{\adg}-1\bigr)\Bigr]^{-1} \ez\Xil \epsilon_{(-1,0)}
\nonumber \\[.5ex]
&= \Xil\ez \epsilon_{(-1,0)} + \biggl( 1- \e^{\adg}\Bigl[1+ \ez\Xil\bigl(\e^{\adg}-1\bigr)\Bigr]^{-1}\biggr) \ez\Xil \epsilon_{(-1,0)}
\nonumber \\[.5ex]
&= \Xil\ez \epsilon_{(-1,0)} - \Xil\ez\bigl(\e^{\adg}-1\bigr)\Bigl[1+ \ez\Xil\bigl(\e^{\adg}-1\bigr)\Bigr]^{-1}\ez\Xil \epsilon_{(-1,0)}\,,
\end{align}
which indeed satisfies \eqref{Xil:transf:2}.
We have thus completed the first part of our proof.

Let us next move on to the second part.
Under the transformation \eqref{del1:app} of the parameters, the gauge transformation 
\begin{equation} \label{del0:G}
\delta_0 G = g\Bigl[ \bigl( Q\hs\epsilon_{(-1,0)} \bigr) G + G\bigl( \ez\hs\epsilon_{(-1,1)}\bigr)\Bigr]
\end{equation}
changes by
\begin{equation} \label{del1:del0:G}
\delta_1\bigl(\delta_0 G\bigr) = g\, G\bigl\{\tQ, \ez\bigr\}\epsilon_{(-2,1)} 
= -\iu\,g^3\hs G\left[\frac{\delta \SB}{\delta G} G\hs,\hs \epsilon_{(-2,1)}\right].
\end{equation}
In the last equality of \eqref{del1:del0:G}, we have used \eqref{tQez:A}.
Because \eqref{del1:del0:G} is nonzero,
the gauge tranformation specified by the original parameters $\epsilon_{(-1,0)}$ and $\epsilon_{(-1,1)}$
and the one specified by their transforms $\epsilon_{(-1,0)} + \delta_1\epsilon_{(-1,0)}$ and $\epsilon_{(-1,1)}+ \delta_1\epsilon_{(-1,1)}$ 
are obviously different.
However, these two gauge transformations are effectively the same because their difference \eqref{del1:del0:G}
is nothing but a trivial gauge transformation.
Indeed, the variation of the action under the transformation
\begin{equation}
G\ \longrightarrow\ G + \delta_1\bigl(\delta_0 G\bigr)
\end{equation}
trivially vanishes as follows:
\begin{align}
\delta \SB &= \left\langle \frac{\delta \SB}{\delta G}\,,\, \delta_1\bigl(\delta_0 G\bigr) \right\rangle
= -\iu\,g^3 \left\langle \frac{\delta \SB}{\delta G} G\,,\, \left[\frac{\delta \SB}{\delta G} G\hs,\hs \epsilon_{(-2,1)}\right] \right\rangle
\nonumber \\*
&= -\iu\,g^3 \left\langle \left[ \frac{\delta \SB}{\delta G} G\hs,\hs \frac{\delta \SB}{\delta G} G\right] ,\,\epsilon_{(-2,1)} \right\rangle
= 0\,.
\end{align}
Eq.~\eqref{del1:del0:G} is different from \eqref{trv} in appearance,
but non-commutativity of the multiplication in string field theory admits trivial gauge transformations of the form \eqref{del1:del0:G}.
If we consider the gauge transformation generated by $\epsilon_{(-1,0)}+\delta_1\epsilon_{(-1,0)}$ and $\epsilon_{(-1,1)}+\delta_1\epsilon_{(-1,1)}$,
and minus the trivial transformation \eqref{del1:del0:G} simultaneously, the resultant transformation is exactly the same as \eqref{del0:G}.
In this sense, the original parameters $\epsilon_{(-1,0)}$ and $\epsilon_{(-1,1)}$ 
and their transforms $\epsilon_{(-1,0)} + \delta_1\epsilon_{(-1,0)}$ and $\epsilon_{(-1,1)}+ \delta_1\epsilon_{(-1,1)}$
provide the same gauge transformation.
Therefore, all the possible genuine gauge transformations can be covered by the $\epsilon_{(-1,0)}$ satisfying \eqref{Xil:LQ:app}
and $\epsilon_{(-1,1)}$ in the large Hilbert space. 
We thus conclude that without loss of generality we may assume \eqref{Xil:LQ:app}.

%====================================================================================================================

\end{document}